\documentclass[aps,prb,twocolumn,groupedaddress,amsmath,amssymb,amsfonts]{revtex4}

\usepackage{latexsym,amssymb}
\usepackage{graphicx,color,pstricks}
\usepackage{epsfig,epsf,bm}

\definecolor{gray}{rgb}{0.7,0.7,0.7}

\newcommand{\abs}[1]{\left\vert #1 \right\vert} 

\begin{document}

\title{On the renormalization of Coulomb interactions in two-dimensional tilted Dirac fermions}

\author{Yu-Wen Lee}
\email{ywlee@thu.edu.tw}
\affiliation{Department of Applied Physics, Tunghai University, Taichung, Taiwan, R.O.C.}

\author{Yu-Li Lee}
\email{yllee@cc.ncue.edu.tw}
\affiliation{Department of Physics, National Changhua University of Education, Changhua, Taiwan, R.O.C.}

\date{\today}

\begin{abstract}
We investigate the effects of long-ranged Coulomb interactions in a tilted Dirac semimetal in two
dimensions by using the perturbative renormalization-group method. Depending on the magnitude of the
tilting parameter, the undoped system can have either Fermi points (type-I) or Fermi lines (type-II).
Previous studies usually performed the renormalization-group transformations by integrating out the
modes with large momenta. This is problematic when the Fermi surface is open, like type-II Dirac
fermions. In this work, we study the effects of Coulomb interactions, following the spirit of
Shankar\cite{Shankar}, by introducing a cutoff in the energy scale around the Fermi surface and
integrating out the high-energy modes. For type-I Dirac fermions, our result is consistent with that of
the previous work. On the other hand, we find that for type-II Dirac fermions, the magnitude of the
tilting parameter increases monotonically with lowering energies. This implies the stability of type-II
Dirac fermions in the presence of Coulomb interactions, in contrast with previous results. Furthermore,
for type-II Dirac fermions, the velocities in different directions acquire different renormalization
even if they have the same bare values. By taking into account the renormalization of the tilting
parameter and the velocities due to the Coulomb interactions, we show that while the presence of a
charged impurity leads only to charge redistribution around the impurity for type-I Dirac fermions, for
type-II Dirac fermions, the impurity charge is completely screened, albeit with a very long screening
length. The latter indicates that the temperature dependence of physical observables are essentially
determined by the RG equations we derived. We illustrate this by calculating the temperature dependence
of the compressibility and specific heat of the interacting tilted Dirac fermions.

\end{abstract}

\maketitle

\section{Introduction}

The Weyl fermions in solid state materials have attracted intense theoretical and experimental interests
in condensed matter community in recent years. These materials are topological since the Weyl nodes, at
which the conduction and the valence bands touch in the momentum space, act as the magnetic monopoles in
the momentum space\cite{XWan}. Therefore, these Weyl nodes can be characterized by the ``magnetic
charges" they carry. On the other hand, many of the electromagnetic responses or the transport
properties of the Weyl semimetal are deeply rooted in nontrivial phenomena in the quantum field theory,
such as the chiral anomaly\cite{Zyuzin,ZWang,CXLiu,PGoswami,Hosur}. Very recently, Weyl fermions have
been detected experimentally in the non-centrosymmetric but time-reversal preserving materials such as
TaAs, NbAs, TaP, and NbP\cite{CShekhar,BQLv,SYXu,BLv,LYang,SYXu2,NXu}.

Due to the lack of a fundamental Lorentz symmetry, the spectra of Dirac/Weyl semimetals realized in
solid state materials do not have to be isotropic. In particular, they can be tilted\cite{AASoluyanov}.
When the tilting angle is large enough, the electron and hole Fermi surfaces can coexist with the
band-touching Dirac/Weyl nodes. This leads to a new kind of materials, which are commonly referred to
as type-II Dirac/Weyl semimetals\cite{AASoluyanov}. In three dimensions ($3$D), the tilted Weyl cones
were proposed to be realized in a material WTe$_2$\cite{AASoluyanov}, while in two dimensions ($2$D),
the tilted Dirac cones were proposed to be realized in a mechanically deformed graphene and the organic
compound $\alpha$-(BEDT-TTF)$_2$I$_3$\cite{MOGoerbig}. Recently, type-II Dirac fermions are
experimentally discovered in two materials: PdTe$_2$\cite{HJNoh,FCFei} and PtTe$_2$\cite{MZYan}.

In contrast with type-I Dirac/Weyl semimetals whose Fermi surface is point-like, type-II Dirac/Weyl
semimetals have an extended Fermi surface. In $2$D, it consists of two straight lines crossing at the
Dirac node, while it is hyperboloids touched at the Weyl nodes in $3$D. Due to the nonvanishing density
of states (DOS) at the Fermi energy, the physical properties of type-II Dirac/Weyl semimetals are
expected to be distinct from those of type-I Dirac/Weyl semimetals. On the other hand, the open Fermi
surface in the type-II materials will result in behaviors different from the usual Fermi liquid (FL)
which has a closed Fermi surface. Even for type-I materials, the nonzero tilting parameter may lead to
observable effects. Some theoretical studies along these directions mainly for non-interacting fermions
have been performed, including the conductance and noise for tilted Dirac ($2$D) or Weyl fermions
($3$D)\cite{MTrescher}, longitudinal magnetoconductivity for tilted Dirac fermions\cite{Pros},
thermodynamic and optical responses for tilted Weyl fermions in the presence of magnetic
fields\cite{TEOBrien,ZMYu,MUdagawa,STchoumakov}, anomalous Hall effect for tilted Weyl
fermions\cite{AAZyuzin,JFSteiner}, and anamolous Nernst and thermal Hall effect for tilted Weyl
fermions\cite{Ferreiros,Saha}.

In the present work, we study the effects of long-range Coulomb interactions on the tilted Dirac
fermions in terms of the renormalization group (RG). Due to the vanishing DOS at the Fermi level, it is
well-known that the Coulomb interaction for untilted Dirac fermions in $2$D is marginally irrelevant in
the sense of RG\cite{JGonzalez,JGonzalez2,DTSon}. Moreover, the value of Fermi velocity will increase
at low enegies. This leads to logarithmic corrections to various thermodynamic response functions,
which can be computed in terms of the RG equations. The results fit the experimental data for graphene
quite well\cite{DESheehy}.

What will happen for Dirac fermions in $2$D when the tilting parameter is not zero? Previous RG
analysis\cite{HIsobe,ZMHuang} shows that the values of velocities also increase logarithmically toward
infinity, as what happens in graphene. By taking into account the fluctuations of transverse gauge
fields, the velocities of fermions are renormalized up to the speed of light at low
energies\cite{HIsobe2}. Moreover, according to the analysis in Refs. \onlinecite{HIsobe2} and
\onlinecite{ZMHuang}, the magnitude of the tilting parameter flows to zero at low energies for both
types of Dirac fermions. For type-II Dirac fermions, this indicates its instability in the presence of
Coulomb interactions. Within such a scenario, interacting type-II Dirac fermions is stable only when
the screening length is short enough so that the above RG flow stops and turns its direction before
the instability occurs.

All the above mentioned RG studies on tilted Dirac fermions employ some types of cutoff functions in
the momenta, such as a direct momentum cutoff or the dimensional regularization. This is certainly fine
for type-I Dirac fermions since the Fermi surface is a point. However, for systems with open Fermi
surface like type-II Dirac fermions, any momentum cutoff function, which is a curve in the momentum
space, will intersect with the Fermi surface. By integrating out the modes with momenta larger than the
mometum curoff to implement the RG transformation, one integrates out not only the high-energy modes
but also the low-energy modes around the Fermi surface. Hence, the RG transformation is not to scale to
the Fermi surface, and the conclusions about type-II Dirac fermions inferring from such a RG analysis
is dubious.

As emphasized by Shankar\cite{Shankar}, the proper way of doing RG for FL should be to scale toward the
Fermi surface. The usual FL has a closed Fermi surface. Therefore, there exists a characteristic
momentum, the Fermi momentum, to separate the high-energy modes from the low-energy modes. For type-II
Dirac fermios, its Fermi surface is open so that such a characteristic momentum simply does not exist.
To implement the RG transformation properly, we label the excitations directly by their energies and
other ''angle" variables. By integrating out the high-energy modes, we obtainn the RG functions for
velocities of fermios and the tilting parameter. Our main findings are as follows.

(i) For type-I Dirac fermions, our results are identical to the previous ones. That is, the values of
velocities and the magnitude of the tilting parameter monotonically increase and decrease at low
energies, respectively. This is not surprising since the Fermi surface is point-like in this case.

(ii) For type-II Dirac fermions, the values of velocities are also increase monotonically at low
energies. In contrast with the previous studies, the RG functions for the two velocities of fermions
are different so that they have different RG flow even if the bare values are the same. Moreover, the
magnitude of the tilting parameter increases as lowering the energy, indicating the stability of
type-II Dirac fermions. In other words, the long-range Coulomb interaction helps to stabilize type-II
Dirac fermions, in contrast with previous studies.

(iii) In terms of the RG equations, we calculate various thermodynamic response functions for both
types of Dirac fermions: the temperature dependence of specific heat at low temperatures, and the
density and temperature dependence of the isothermal compressibility.

(iv) Since there is a nonzero DOS at the Fermi level for type-II Dirac fermions, it is argued that the
Coulomb interaction will be screened at long distances.\cite{HIsobe2, ZMHuang} In order to address this
issue, we calculate the vacuum polarization at zero frequency. We find that the polarization function
is singular at zero momentum for both types of Dirac fermions. This singularity can be removed if we
sum the diagrams with leading divergences in terms of RG, following the method employed in Ref.
\onlinecite{RRBiswas}. From this RG-improved polarization function, we find that there is indeed
complete screening for type-II Dirac fermions, albeit with a very long screening length. This indicates
that the scaling of physical observables with temperatures in a large temperature range is indeed
described by the RG equations we derived in this paper.

The organization of the rest of the paper is as follows. The model is defined and discussed in Sec.
\ref{model}. We present the RG equations and its implications in Sec. \ref{rge}. The calcutations of
thermodynamic response functions for both types of Dirac fermions are shown in Sec. \ref{thermo}. Sec.
\ref{coul} is about the screening of the Coulomb potential and the Coulomb impurity problem. The last
section is devoted to conclusive discussions. The details of calculations of the RG equations and the
vacuum polarization are put in appendix \ref{self} and \ref{pol}, respectively.

\section{The model}
\label{model}

We start with the minimal model of non-interacting tilted Dirac fermions described by the Hamiltonian\cite{MOGoerbig}
\begin{equation}
 H_0=\! \sum_{\xi,\sigma,\bm{p}}\tilde{\psi}_{\xi\sigma}^{\dagger}(\bm{p})(\xi wv_1p_1+\xi v_1p_1\sigma_1+v_2p_2)
 \tilde{\psi}_{\xi\sigma}(\bm{p}) \ , \label{wf2h1}
\end{equation}
where $\xi=\pm 1$ denote the valley degeneracy, $\sigma=\pm 1$ account for the spin degeneracy, and
$\sigma_{1,2,3}$ are the standard Pauli matrices. The fields $\tilde{\psi}_{\xi\sigma}(\bm{p})$ and
$\tilde{\psi}^{\dagger}_{\xi\sigma}(\bm{p})$, which obey the canonical anticommutation relations,
describe Dirac fermions around the Dirac nodes at the points $\bm{K}$ and $-\bm{K}$ in the first
Brillouin zone (BZ). $v_1$ and $v_2$ are the ``speeds" of Dirac fermions along the $x$ and $y$
directions, respectively. Without loss of generality, we take $v_1,v_2>0$. In Eq. (\ref{wf2h1}), we
have set the energy of the Dirac nodes to be zero. The dimensionless constant $w$ is the tilting
parameter.

\begin{figure}
\begin{center}
 \includegraphics[width=0.95\columnwidth]{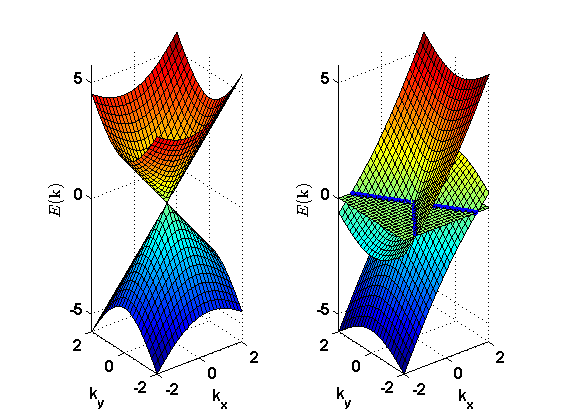}
 \caption{Conical band structure ($\xi=1$) for type-I Dirac fermions (left) and type-II Dirac fermions
 (right). For type-II Dirac fermions, the blue lines on the $E=0$ plane indicate the Fermi surface.}
 \label{wf2f1}
\end{center}
\end{figure}

The spectrum of $H_0$ is given by
\begin{equation}
 \epsilon_{\pm}(\bm{p})=\xi w\tilde{p}_1\pm\sqrt{\tilde{p}_1^2+\tilde{p}_2^2} \ , \label{wf2h11}
\end{equation}
where $\tilde{p}_i=v_ip_i$ with $i=1,2$. It is clear that the Dirac cones are tilted along the $x$-axis
when $w\neq 0$, as shown in Fig. \ref{wf2f1}. When $|w|<1$, the Dirac cones are slightly tilted,
corresponding to type-I Dirac fermions, while the Dirac cones are tipped over when $|w|>1$,
corresponding to type-II Dirac fermions. $|w|=1$ are the Lifshitz transition points which separate the
two types of Dirac fermions. When the chemical potential $\mu=0$, the Fermi surface consists of two
straight lines:
\begin{equation}
\tilde{p}_2=\pm\tilde{w}\tilde{p}_1 \ , \label{wf2h12}
\end{equation}
for type-II Dirac fermions, where $\tilde{w}=\sqrt{w^2-1}$.

The Coulomb interaction between electrons is described by the Hamiltonian
\begin{equation}
 H_C=\frac{1}{2} \! \int \! d^2xd^2y\rho(\bm{x})V(|\bm{x}-\bm{y}|)\rho(\bm{y}) \ , \label{wf2h13}
\end{equation}
where $\rho(\bm{r})=\sum_{\sigma=\pm}:c^{\dagger}_{\alpha}(\bm{r})c_{\sigma}(\bm{r}):$ is the
normal-ordered electron density operator with the annihilation operator $c_{\sigma}$ and creation
operator $c^{\dagger}_{\sigma}$ for electrons with spin-$\sigma$, and
\begin{equation}
 V_0(\bm{r})=\frac{e^2}{4\pi\epsilon r} \ , \label{wf2h14}
\end{equation}
is the Coulomb potential. In Eq. (\ref{wf2h14}), $\epsilon$ is the dielectric constant and $-e$ is the
charge carried by the electron. In terms of the low-energy degrees of freedom around the Dirac nodes,
the electron operator $c_{\sigma}$ can be written as
\begin{equation}
 c_{\sigma}(\bm{r})=\! \sum_{\xi=\pm 1}e^{i\xi\bm{K}\cdot\bm{r}}\psi_{\xi\sigma}(\bm{r})+\cdots \ ,
 \label{wf2low1}
\end{equation}
where the fermion field $\psi_{\xi\sigma}(\bm{r})=\frac{1}{\sqrt{A}}\sum_{\bm{p}}e^{i\bm{p}\cdot\bm{r}}
\tilde{\psi}_{\xi\sigma}(\bm{p})$ and $A$ is the area of the system. With the help of Eq.
(\ref{wf2low1}), the density operator can be written as
\begin{equation}
 \rho(\bm{r}) \! =\rho_0(\bm{r})+ \! \left[e^{-2i\bm{K}\cdot\bm{r}}M(\bm{r})+\mathrm{H.c.}\right] \! +
 \cdots \ , \label{wf2low11}
\end{equation}
where $\rho_0(\bm{r})=\! \sum_{\xi,\sigma}:\psi^{\dagger}_{\xi\sigma}\psi_{\xi\sigma}(\bm{r}):$
describes the uniform component of $\rho$, while
$M(\bm{r})=\! \sum_{\sigma}\psi^{\dagger}_{+\sigma}\psi_{-\sigma}(\bm{r})$ is the order parameter for
the charge-density-wave (CDW) ordering. Substituting Eq. (\ref{wf2low11}) into $H_c$ [Eq.
(\ref{wf2h13})], we find that $H_c=H_{int}+\cdots$, where
\begin{equation}
 H_{int}=\frac{1}{2} \! \int \! d^2xd^2y\rho_0(\bm{x})V_0(|\bm{x}-\bm{y}|)\rho_0(\bm{y}) \ ,
 \label{wf2h15}
\end{equation}
and $\cdots$ contains those terms with the factors $e^{\pm 2i\bm{K}\cdot\bm{r}}$ or
$e^{\pm 2i\bm{K}\cdot(\bm{x}\pm\bm{y})}$. Due to the fast oscillating nature, these terms will
generate short-ranged repulsive four-fermion interactions at low energies, and we shall neglect them.

Our working Hamiltonian is $H_0+H_{int}$. When $\mu=0$, $H$ is invariant against the ``particle-hole"
(PH) transformation
\begin{equation}
 \tilde{\psi}_{\xi\sigma}(\bm{p})\rightarrow\sigma_1\tilde{\psi}^*_{\xi\sigma}(-\bm{p}) \ .
 \label{wf2ph1}
\end{equation}
This PH symmetry forbids terms like $\tilde{\psi}_{\xi\sigma}^{\dagger}\tilde{\psi}_{\xi\sigma}$,
$\tilde{\psi}_{\xi\sigma}^{\dagger}\sigma_1\tilde{\psi}_{\xi\sigma}$, or
$\tilde{\psi}_{\xi\sigma}^{\dagger}\sigma_2\tilde{\psi}_{\xi\sigma}$ since they are odd under the PH
transformation. We shall see later that this PH symmetry together with gauge invariance guarantee the
renormalizablity of this theory.

\section{The RG equations}
\label{rge}

To derive the RG equation, we perform a Hubbard-Stratonovich transformation so that the action in
the imaginary-time formulation can be written as
\begin{eqnarray}
 S \! \! &=& \! \! \! \sum_{\xi,\alpha} \! \! \int_X\psi^{\dagger}_{\xi\alpha}
 [\partial_{\tau}-i\xi v_1(w+\sigma_1)\partial_1-iv_2\sigma_2\partial_2]\psi_{\xi\alpha} \nonumber \\
 \! \! & & \! \! +\! \int_Q\frac{|\bm{q}|}{g^2}\tilde{\phi}^{\dagger}(Q)\tilde{\phi}(Q)+i \! \! \int_X
 \phi(X)\rho_0(X) \ , \label{wf2s2}
\end{eqnarray}
where $Q=(iq_0,\bm{q})$, $X=(\tau,\bm{x})$, $g^2=e^2/\epsilon$, $\int_X=\! \int \! d\tau d^2x$,
$\int_Q=\! \int \! \frac{dq_0}{2\pi}\frac{d^2q}{(2\pi)^2}$, and $\tilde{A}(Q)$ denotes the Fourier
transform of $A(X)$. Since the auxillary field $\phi(X)$ is real,
$\tilde{\phi}^{\dagger}(Q)=\tilde{\phi}(-Q)$. We have extended the number of fermion fields to $N$
pairs such that $\alpha=1,2,\cdots,N$. Physically, $N=2$ due to the spin degeneracy. $S$ still preserves
the PH symmetry as long as we require that $\phi$ transforms as
\begin{equation}
 \phi\rightarrow -\phi \ , \label{wf2ph11}
\end{equation}
under the PH transformation.

As we have discussed in the introduction, the proper way to implement the RG transformation for a
system with an open Fermi surface is to integrate out an energy shell each time, instead of a momentum
shell. To achieve this gola, we have to parametrize the euqla-energy curves first. These curves are
given by the equations $\epsilon_{\pm}(\bm{p})=E$. For type-I Dirac fermions ($|w|<1$), these
equal-energy curves are ellipses and can be parametrized as
\begin{eqnarray}
\tilde{p}_1 &=& -\frac{\xi w}{1-w^2}E+\frac{|E|}{1-w^2}\cos{\theta} \ , \nonumber \\
\tilde{p}_2 &=& \frac{|E|}{\sqrt{1-w^2}}\sin{\theta} \ , \label{wf2e1}
\end{eqnarray}
for given $E$, where $0\leq\theta<2\pi$. On the other hand, for type-II Dirac fermions ($|w|>1$), these
equal-energy curves are hyperbolas and can be parametrized as
\begin{eqnarray}
\tilde{p}_1 &=& \frac{\xi w}{w^2-1}E\pm\frac{|E|}{w^2-1}\cosh{\theta} \ , \nonumber \\
\tilde{p}_2 &=& \frac{|E|}{\sqrt{w^2-1}}\sinh{\theta} \ , \label{wf2e11}
\end{eqnarray}
for given $E$, where $-\infty<\theta<+\infty$. In Eq. (\ref{wf2e11}), the $+$ and $-$ signs correspond
to the right and the left branches of the hyperbola, respectively.

In terms of the parametrization (\ref{wf2e1}) or (\ref{wf2e11}), the momentum integral can be written as
\begin{equation}
 \int \! d^2\tilde{p}=\frac{1}{2} \! \int_{-\Lambda}^{\Lambda} \! \frac{|E|dE}{(1-w^2)^{3/2}} \!
 \int^{2\pi}_0 \! \! d\theta(1-\eta_E\xi w\cos{\theta}) \ , \label{wf2e12}
\end{equation}
for type-I Dirac fermions, and
\begin{eqnarray}
 \int \! d^2\tilde{p} &=& \frac{1}{2} \! \int_{-\Lambda}^{\Lambda} \! \frac{|E|dE}{\tilde{w}^3} \!
 \left[\! \int^{+\infty}_{-\infty} \! d\theta(|w|\cosh{\theta}+\eta_E\eta_w\xi)\right. \nonumber \\
 & & +\! \left. \! \int^{+\infty}_{-\infty} \! d\theta(|w|\cosh{\theta}-\eta_E\eta_w\xi)\right] ,
 \label{wf2e13}
\end{eqnarray}
for type-II Dirac fermions, where $\Lambda$ is the UV cutoff in energies and $\eta_A=\mbox{sgn}A$. In
Eq. (\ref{wf2e13}), the first and the second $\theta$ integrals for given $E$ correspond to the
integrations over the right and the left branches of the hyperbola, respectively. In fact, it suffices
to consider the integrals over $E>0$ or $E<0$ since the involved two bands have been taken into account
by the Pauli matrices. However, this regularization breaks the PH symmetry at $\mu=0$. Hence, we define
the momentum integral by Eqs. (\ref{wf2e12}) or (\ref{wf2e13}), and include a prefactor $1/2$.

Before plunging into the calculation of the RG equations, we discuss some constraints on the
renormalization of the various terms in the action $S$. First of all, $S$ is invariant against the
gauge transformation:
\begin{equation}
 \psi_{\xi\alpha}\rightarrow e^{-i\chi(\tau)}\psi_{\xi\alpha} \ ,
 ~\phi\rightarrow\phi+\partial_{\tau}\chi \ . \label{wf2wi1}
\end{equation}
By integrating out the fast modes which have energies within the range $(\Lambda/s,\Lambda)$ with
$s=e^l>1$, $S$ becomes
\begin{eqnarray*}
 S &\rightarrow& (1+\Sigma_{\tau}) \! \sum_{\xi,\alpha} \! \int_X\psi^{\dagger}_{\xi\alpha}
 \partial_{\tau}\psi_{\xi\alpha}+(1+\Gamma_0)i \! \int_X\phi\rho_0 \\
 & & +\! \sum_{\xi,\alpha} \! \int_X\psi^{\dagger}_{\xi\alpha}\mathcal{H}_0\psi_{\xi\alpha}+ \! \int_Q
 \frac{|\bm{q}|}{g^2}\tilde{\phi}^{\dagger}(Q)\tilde{\phi}(Q)+\cdots \ ,
\end{eqnarray*}
where $\cdots$ denotes the terms with higher scaling dimensions. Here we do not explicitly write down
the renormalized Hamiltonian $\mathcal{H}_0$ since it is irrelevant in the derivation of the Ward
identity. The gauge invariance leads to the Ward identity\cite{JGonzalez,GYCho}
\begin{equation}
 \Sigma_{\tau}=\Gamma_0 \ . \label{wf2wi11}
\end{equation}
Next, the non-analytic dependence of the $|\bm{q}|/g^2$ term on $\bm{q}$ prohibits its renormalization
under integrating out the fast modes\cite{JGonzalez,GYCho}. Thus, it is not necessary to introduce the
wavefunction renormalization of the $\phi$ field. This fact together with Eq. (\ref{wf2wi11}) result
in the non-renormalization of $g^2$.

We now integrate out the fast modes to the one-loop order to get the RG equations by assuming the
weak-coupling limit $g^2/\sqrt{v_1v_2}\ll 1$. From the above discussions, it suffices to compute the
self-energy of fermions. After integrating out the fast modes and rescaling the energy, angle variable,
frequency, and fields by $E\rightarrow s^{-1}E$, $\theta\rightarrow\theta$, $p_0\rightarrow s^{-1}p_0$,
$\tilde{\psi}_{\xi\alpha}\rightarrow s^2Z^{-1/2}_{\psi}\tilde{\psi}_{\xi\alpha}$, and
$\tilde{\phi}\rightarrow s^2\tilde{\phi}$, we find the following facts: (i)
$\Sigma_{\tau}=O(g^4/v_1v_2)$ to the one-loop order, which leads to $Z_{\psi}=1+O(g^4/v_1v_2)$ for both
types of Dirac fermions. (ii) $v_1$ and $v_2$ acquire nontrivial renormalization. (iii) $wv_1$ is not
renormalized to the one-loop order for type-I Dirac fermions, while it acquires nontrivial
renormalization for type-II Dirac fermions. The details of the claculations are left to appendix
\ref{self}.

The resulting one-loop RG equations for type-I Dirac fermions are
\begin{equation}
 \frac{dv_l}{dl}=\frac{g^2}{16\pi} \ , ~~
 \frac{d}{dl}(w_lv_l)=0 \ , \label{wf2rge2}
\end{equation}
where the quantities with the subscript $l$ indicate the renormalized parameters, while those without
the subscript $l$ correspond to the bare ones. Here we have taken $v_1=v_2=v$. Equation (\ref{wf2rge2})
is identical to the one in Ref. \onlinecite{HIsobe}. On the other hand, the one-loop RG equations for
type-II Dirac fermions are
\begin{eqnarray}
 \frac{dv_{1l}}{dl} &=& \frac{\tilde{w}_lg^2r_l}{8\pi^2}M_1(|w_l|,r_l) \ , \label{wf2rge21} \\
 \frac{dv_{2l}}{dl} &=& \frac{\tilde{w}_l^3g^2r_l^2}{8\pi^2}M_2(|w_l|,r_l) \ , \label{wf2rge22}
\end{eqnarray}
and
\begin{equation}
 \frac{d|w_l|}{dl}=\frac{\tilde{w}_lg^2r_l}{8\pi^2v_{1l}}\mathcal{N}(|w_l|,r_l) \ , \label{wf2rge23}
\end{equation}
where $r_l=v_{1l}/v_{2l}$,
\begin{eqnarray*}
 M_1 &=& \! \int^{+\infty}_{\ln{|w|}} \! d\theta\frac{(|w|-\cosh{\theta})^2}
 {[(|w|-\cosh{\theta})^2+r^2\tilde{w}^2\sinh^2{\theta}]^{3/2}} \ , \\
 M_2 &=& \! \int^{+\infty}_{\ln{|w|}} \! d\theta\frac{\sinh^2{\theta}}
 {[(|w|-\cosh{\theta})^2+r^2\tilde{w}^2\sinh^2{\theta}]^{3/2}} \ ,
\end{eqnarray*}
and
\begin{eqnarray*}
 \mathcal{N} &=& \! \int^{+\infty}_0 \! d\theta\frac{(|w|\cosh{\theta}+1)(|w|+\cosh{\theta})}
 {[(|w|+\cosh{\theta})^2+r^2\tilde{w}^2\sinh^2{\theta}]^{3/2}} \\
 & & + \! \int^{+\infty}_0 \! d\theta\frac{(|w|\cosh{\theta}-1)(|w|-\cosh{\theta})}
 {[(|w|-\cosh{\theta})^2+r^2\tilde{w}^2\sinh^2{\theta}]^{3/2}} \\
 & & -\! \int^{+\infty}_{\ln{|w|}} \! d\theta\frac{\tilde{w}^2(|w|-\cosh{\theta})}
 {[(|w|-\cosh{\theta})^2+r^2\tilde{w}^2\sinh^2{\theta}]^{3/2}} \ .
\end{eqnarray*}
Since $M_1,M_2,r>0$, Eqs. (\ref{wf2rge21}) and (\ref{wf2rge22}) indicate that the values of $v_{1l}$
and $v_{2l}$ become large at low energies. This implies that the dimensionless coupling
$\lambda=Ng^2/(16\sqrt{v_1v_2})$ is irrelevant around the Gaussian (non-interacting) fixed point, and
thus justifies our perturbative calculations.

\begin{figure}
\begin{center}
 \includegraphics[width=0.95\columnwidth]{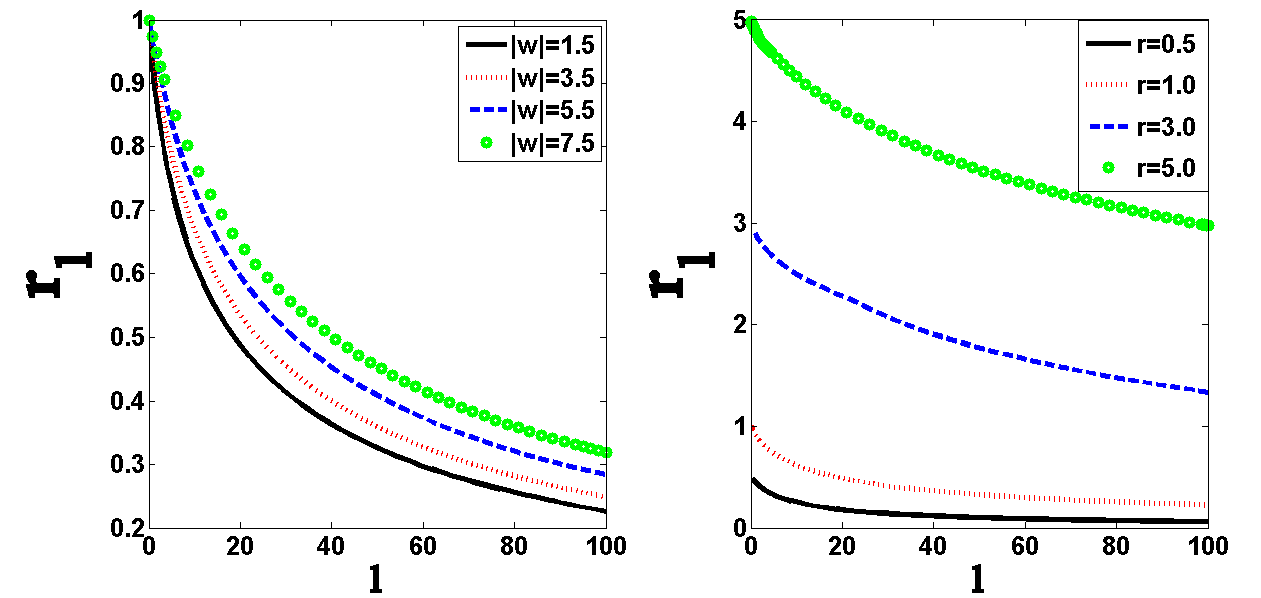}
 \caption{The RG flow of the ratio $r_l=v_{1l}/v_{2l}$ with $r(0)=1$ and various values of $|w|$ (left)
 and with $|w|=1.5$ and various various values of $r$ (right) for type-II Dirac fermions. In both
 diagrams, we have taken $v_2=1.3g^2/(4\pi)$.}
 \label{wf2rgef1}
\end{center}
\end{figure}

\begin{figure}
\begin{center}
 \includegraphics[width=0.95\columnwidth]{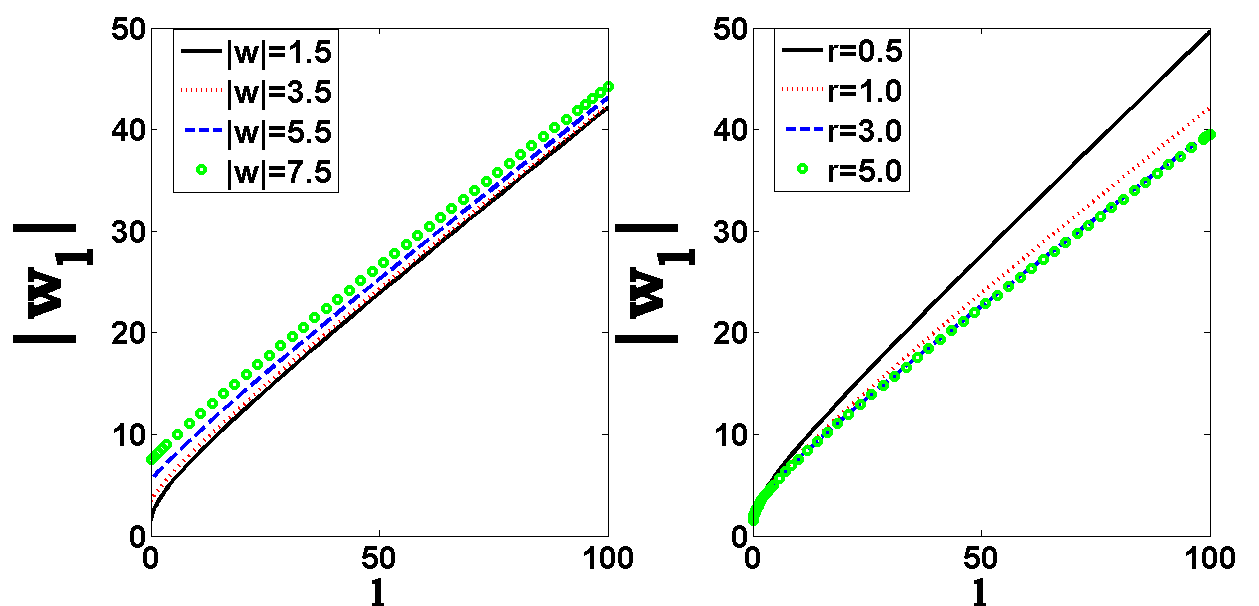}
 \caption{The RG flow of $|w_l|$ with $r=1$ and various values of $|w|$ (left) and with $|w|=1.5$ and
 various values of $r$ (right) for type-II Dirac fermions. In both diagrams, we have taken
 $v_2=1.3g^2/(4\pi)$.}
 \label{wf2rgef11}
\end{center}
\end{figure}

The RG flows for the ratio $r_l$ and the tilting parameter $|w_l|$ are shown in Figs. \ref{wf2rgef1} and
\ref{wf2rgef11}, respectively. We see that although both $v_{1l}$ and $v_{2l}$ increase at low energies,
the rate of change of $v_{2l}$ is faster than that of $v_{1l}$ so that the ratio $r_l$ decreases at low
energies. Therefore, even if we take $v_1=v_2$, $r_l\neq 1$ at $l>0$. This is different from type-I
Dirac fermions where $r_l=1$ if $v_1=v_2$. On the other hand, the value of $|w_l|$ always increases at
low energies. This implies the stability of type-II Dirac fermions against weak Coulomb repulsions. Our
result is distinct from the conclusion in Ref. \onlinecite{ZMHuang}.

\section{Thermodynamics at low temperaures or small densities}
\label{thermo}

Now we are in a position to extract the temperature or density dependence of various thermodynamic
response functions at low temperatures or small densities with the help of the RG equations, following
the method proposed in Ref. \onlinecite{DESheehy}.

Before doing it, we have to determine the RG flows of the temperature $T$ and the chemical potential
$\mu$. At finite temperature, we find that
\begin{eqnarray*}
 \tau\rightarrow s\tau \ ,
\end{eqnarray*}
according to the rescaling of $p_0$. Hence, we get $(T^{\prime})^{-1}=s^{-1}T^{-1}$ or
\begin{equation}
\frac{dT_l}{dl}=T_l \ . \label{wf2rge24}
\end{equation}
The solution of Eq. (\ref{wf2rge24}) is $T_l=Te^l$.

To extract the scaling equation for $\mu$, we add the term
\begin{eqnarray*}
 -\mu \! \int_X \! \rho(X)=-\mu \! \sum_{\xi,\sigma} \! \int_P \! \tilde{\psi}^{\dagger}_{\xi\sigma}
 (P)\tilde{\psi}_{\xi\sigma}(P) \ ,
\end{eqnarray*}
to the action. By integrating out the fast modes, this term becomes
\begin{eqnarray*}
 -\frac{\mu}{v_1v_2} \! \sum_{\xi,\sigma} \! \int^{+\infty}_{-\infty} \! \frac{dp_0}{2\pi} \!
 \int_{\Lambda/s} \! \frac{d^2\tilde{p}}{(2\pi)^2}\tilde{\psi}^{\dagger}_{\xi\sigma}(P)
 \tilde{\psi}_{\xi\sigma}(P) \ ,
\end{eqnarray*}
to the one-loop order. Performing the rescaling of $E$, $\theta$, $p_0$, and
$\tilde{\psi}_{\xi\sigma}$, we find that
\begin{eqnarray*}
 & & -\frac{\mu}{v_1v_2} \! \sum_{\xi,\sigma} \! \int^{+\infty}_{-\infty} \! \frac{dp_0}{2\pi} \!
	 \int_{\Lambda/s} \! \frac{d^2\tilde{p}}{(2\pi)^2}\tilde{\psi}^{\dagger}_{\xi\sigma}(P)
	 \tilde{\psi}_{\xi\sigma}(P) \\
 & & =-\frac{s\mu}{v_1v_2} \! \sum_{\xi,\sigma} \! \int^{+\infty}_{-\infty} \! \frac{dp_0}{2\pi} \!
	 \int_{\Lambda} \! \frac{d^2\tilde{p}}{(2\pi)^2}\tilde{\psi}^{\dagger}_{\xi\sigma}(P)
	\tilde{\psi}_{\xi\sigma}(P) \\
 & & =-s\mu \! \sum_{\xi,\sigma} \! \int_P \! \tilde{\psi}^{\dagger}_{\xi\sigma}(P)
	 \tilde{\psi}_{\xi\sigma}(P) \ .
\end{eqnarray*}
That is, $\mu^{\prime}=s\mu$ or
\begin{equation}
\frac{d\mu_l}{dl}=\mu_l \ . \label{wf2rge25}
\end{equation}
The solution of Eq. (\ref{wf2rge25}) is $\mu_l=\mu e^l$.

The first physical quantity we would like to study is the isothermal compressibility $\kappa$, which is
defined as $\kappa=(\partial n/\partial\mu)_T$ where $n$ is the average density of electrons. Since $n$
is a physical quantity, upon renormalization, we find that
\begin{equation}
 n(T,\mu,v_1,v_2,w,g^2)=s^{-2}n(T_l,\mu_l,v_{1l},v_{2l},w_l,g^2) \ , \label{wf2phys1}
\end{equation}
where $s=e^l$. We may regard $n(T_l,\mu_l,v_{1l},v_{2l},w_l,g^2)$ as the average density of the
renormalized system where the effective coupling $\lambda_l$ is small and the effective temperature
$T_l$ is high. With an appropriated choice for the renormalization scale $l$, we can put the
renormalized theory into a regime in which the calculation becomes simple.

From Eq. (\ref{wf2phys1}) and the solutions of Eqs. (\ref{wf2rge24}) and (\ref{wf2rge25}), we find that
$\kappa=s^{-1}\kappa_l$. We first consider the case with $\mu=0$, and run the RG flow to the scale
$l=l_*$ such that $T_{l_*}=D$ where $D$ is the bandwidth which is the UV cutoff in energies for the
model we use to describe the Dirac fermions. From the solution of Eq. (\ref{wf2rge24}), we have
$l_*=\ln{(D/T)}$. In terms of this expression, we get $\kappa=e^{-l_*}\kappa_*$ where $A_*=A(l_*)$
denotes the renormalized variable at scale $l=l_*$. Since the effective coupling $\lambda_l$ is
irrelevant, we may replace $k_*$ by the result of non-interacting fermions.

For type-I Dirac fermions, we have
\begin{eqnarray*}
 \kappa_*(\mu=0) &=& \frac{(4\ln{2})T_*}{\pi v_*^2(1-w_*^2)^{3/2}} \\
 &=& \frac{(4\ln{2})D[1+\frac{\lambda}{4}\ln{(D/T)}]^{-2}}
 {\pi v^2\{1-w^2/[1+\frac{\lambda}{4}\ln{(D/T)}]^2\}^{3/2}} \ .
\end{eqnarray*}
As a result, we get
\begin{equation}
 \kappa(T)=\frac{(4\ln{2})T[1+\frac{\lambda}{4}\ln{(D/T)}]^{-2}}
 {\pi v^2\{1-w^2/[1+\frac{\lambda}{4}\ln{(D/T)}]^2\}^{3/2}} \ . \label{wf2phys11}
\end{equation}
Equation (\ref{wf2phys11}) reduces to the one for graphene when $w=0$\cite{DESheehy}. It indicates
that the compressibility at given temperature is enhanced due to a nonzero tilting parameter. On the
other hand, for type-II Dirac fermions, we have
\begin{eqnarray*}
 \kappa_*(\mu=0)=\frac{De^{l_*}}{\pi^2v_{1*}v_{2*}\tilde{w}_*} \ ,
\end{eqnarray*}
leading to
\begin{equation}
 \kappa(T)=\frac{D}{\pi^2v_{1*}v_{2*}\tilde{w}_*} \ . \label{wf2phys12}
\end{equation}

\begin{figure}
\begin{center}
 \includegraphics[width=0.95\columnwidth]{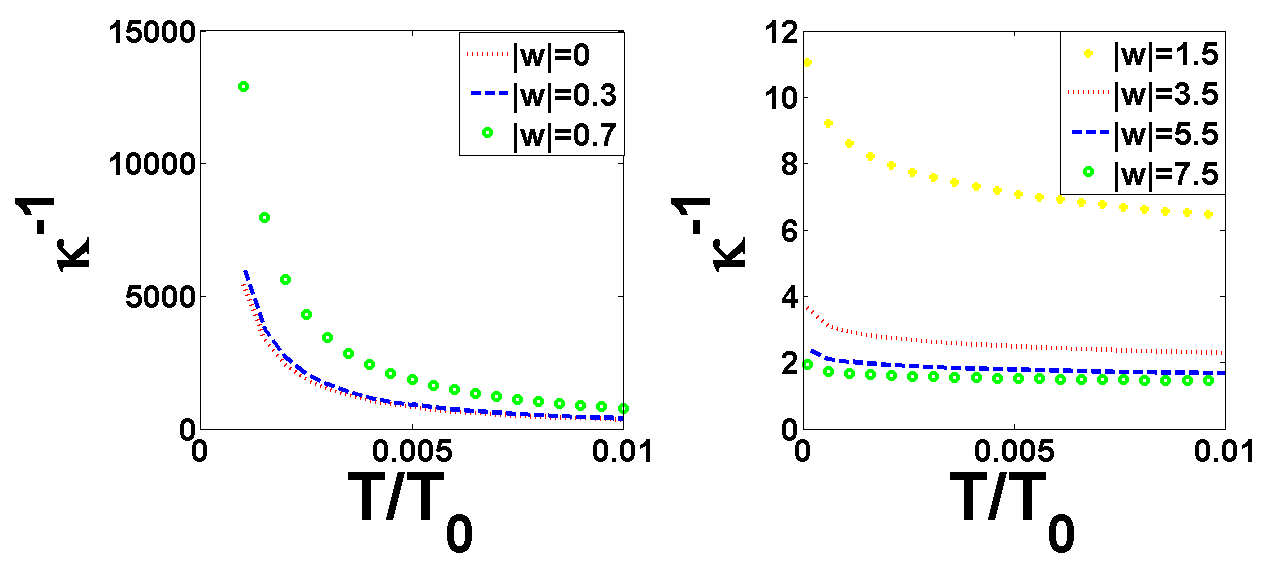}
 \caption{$\kappa^{-1}=(\partial\mu/\partial n)_T$ at $\mu=0$, in units of $\kappa^{-1}_0$, as a
 function of $T/T_0$ for type-I Dirac fermions (left) and type-II Dirac fermions (right), where
 $\kappa_0$ is the compressibility at $T=T_0=D/k_B$. In both diagrams, we take
 $v_1=v_2=1.3g^2/(4\pi)$.}
 \label{wf2thermof1}
\end{center}
\end{figure}

Figure \ref{wf2thermof1} exhibits temperature dependence of $\kappa^{-1}$ in units of $\kappa^{-1}_0$
for both types of Dirac fermions, i.e.,
\begin{eqnarray*}
 \frac{\kappa}{\kappa_0}=\frac{T/T_0}{[1 \! +\! \frac{\lambda}{4}\ln{(T_0/T)}]^2} \! \left\{\frac{1-w^2}
 {1 \! -\! w^2/[1 \! + \! \frac{\lambda}{4}\ln{(T_0/T)}]^2}\right\}^{\! \frac{3}{2}} ,
\end{eqnarray*}
for type-I Dirac fermions, and
\begin{eqnarray*}
 \frac{\kappa}{\kappa_0}=\frac{v_1v_2\tilde{w}}{v_{1*}v_{2*}\tilde{w}_*} \ ,
\end{eqnarray*}
for type-II Dirac fermions, where
\begin{eqnarray*}
 \kappa_0= \! \left\{\begin{array}{cc}
 \frac{(4\ln{2})T_0}{\pi v^2(1-w^2)^{3/2}} & \mbox{type-I} \\
 & \\
 \frac{D}{\pi^2v_1v_2\tilde{w}} & \mbox{type-II}
 \end{array}\right. ,
\end{eqnarray*}
is the compressibility at $T=T_0=D/k_B$. We see that for both types of Dirac fermions, the
compressibility at low temperatures is suppressed by the Coulomb interaction through the enhancement of
the velocities. In particular, the compressibility is a constant for non-interacting type-II Dirac
fermions. Hence, the behavior of $\kappa(T)$ at low temperatures deviates from the one of
non-interacting fermions indicates the effect of Coulomb interactions.

Next, we consider the case with $T=0$, and run the RG flow to the scale $l=l_*$ such that $|\mu_*|=D$,
leading to $l_*=\ln{(D/|\mu|)}$. For type-I Dirac fermions, we have
\begin{eqnarray*}
 n_*(T=0) &=& \frac{|\mu_*|\mu_*}{\pi v_*^2(1-w_*^2)^{3/2}} \\
 &=& \frac{\mbox{sgn}(\mu)D^2[1+\frac{\lambda}{4}\ln{(D/|\mu|)}]^{-2}}
 {\pi v^2\{1-w^2/[1+\frac{\lambda}{4}\ln{(D/|\mu|)}]^2\}^{3/2}} \ ,
\end{eqnarray*}
which leads to
\begin{equation}
 n(\mu)=\frac{|\mu|\mu[1+\frac{\lambda}{4}\ln{(D/|\mu|)}]^{-2}}
 {\pi v^2\{1-w^2/[1+\frac{\lambda}{4}\ln{(D/|\mu|)}]^2\}^{3/2}} \ , \label{wf2phys13}
\end{equation}
and $\kappa(\mu)\approx 2n(\mu)/\mu$. From Eq. (\ref{wf2phys13}), we find that
\begin{eqnarray*}
 |\mu| &\approx& \sqrt{\pi v^2|n|} \! \left[1+\frac{\lambda}{8}\ln{\! \left(\frac{n_0}{|n|}\right)}
 \right] \\
 & & \times \! \left\{1-\frac{w^2}{[1+\frac{\lambda}{8}\ln{(n_0/|n|)}]^2}\right\}^{\! 3/4} ,
\end{eqnarray*}
where $n_0=D^2/[\pi v^2(1-w^2)^{3/2}]$, and thus we obtain
\begin{equation}
 \kappa(n)=\frac{2\sqrt{|n|/\pi}[1+\frac{\lambda}{8}\ln{(n_0/|n|)}]^{-1}}
 {v\{1-w^2/[1+\frac{\lambda}{8}\ln{(n_0/|n|)}]^2\}^{3/4}} \ . \label{wf2phys14}
\end{equation}
When $w=0$, Eqs. (\ref{wf2phys13}) and (\ref{wf2phys14}) reduce to those for graphene\cite{DESheehy}.

On the other hand, for type-II Dirac fermions, we have
\begin{eqnarray*}
 & & n_*(T=0)=\frac{De^{l_*}\mu_*}{\pi^2v_{1*}v_{2*}\tilde{w}_*}=\frac{D^3/\mu}
	 {\pi^2v_{1*}v_{2*}\tilde{w}_*} \ , \\
 & & \kappa_*(T=0)=\frac{De^{l_*}}{\pi^2v_{1*}v_{2*}\tilde{w}_*}=\frac{D^2/|\mu|}
	 {\pi^2v_{1*}v_{2*}\tilde{w}_*} \ .
\end{eqnarray*}
Hence, we get
\begin{equation}
 n(\mu)=\frac{D\mu}{\pi^2v_{1*}v_{2*}\tilde{w}_*} \ , \label{wf2phys15}
\end{equation}
and
\begin{eqnarray*}
 \kappa(\mu)=\frac{D}{\pi^2v_{1*}v_{2*}\tilde{w}_*} \ .
\end{eqnarray*}
From Eq. (\ref{wf2phys15}), we find that $|\mu|=\pi^2v_1v_2\tilde{w}|n|/D$. Substituting this expression
into $\kappa(\mu)$, we obtain
\begin{equation}
 \kappa(n)=\frac{D}{\pi^2v_{1*}v_{2*}\tilde{w}_*} \ , \label{wf2phys16}
\end{equation}
where $l_*=\ln{(n_0/|n|)}$ and $n_0=D^2/(\pi^2v_1v_2\tilde{w})$.

Now we determine the temperature dependence of the specific heat at constant volume, which can be
extracted from the the free energy density $f(T,\mu,v_1,v_2,w,g^2)$ through the relation
$c(T)=-T\partial^2f/\partial T^2$. Upon renormalization, we find that
\begin{equation}
 f(T,\mu,v_1,v_2,w,g^2)=s^{-3}f(T_l,\mu_l,v_{1l},v_{2l},w_l,g^2) \ . \label{wf2phys2}
\end{equation}
Hence, we get $c(T)=e^{-2l_*}c_*$ where $l_*=\ln{(D/T)}$.

For type-I Dirac fermions, we have
\begin{eqnarray*}
 c_*=\frac{18\zeta(3)T_*^2}{\pi v_*^2(1-w_*^2)^{3/2}}=\frac{18\zeta(3)D^2}{\pi v_*^2(1-w_*^2)^{3/2}}
 \ ,
\end{eqnarray*}
which leads to
\begin{equation}
 c(T)=\frac{18\zeta(3)T^2[1+\frac{\lambda}{4}\ln{(D/T)}]^{-2}}
 {\pi v^2\{1-w^2/[1+\frac{\lambda}{4}\ln{(D/T)}]^2\}^{3/2}} \ . \label{wf2phys21}
\end{equation}
Equation (\ref{wf2phys21}) reduces to the one for graphene when $w=0$\cite{DESheehy}. On the other
hand, for type-II Dirac fermions, we have
\begin{eqnarray*}
 c_*=\frac{De^{l_*}T_*}{2v_{1*}v_{2*}\tilde{w}_*}=\frac{D^3/T}{2v_{1*}v_{2*}\tilde{w}_*} \ ,
\end{eqnarray*}
which results in
\begin{equation}
 c(T)=\frac{DT}{2v_{1*}v_{2*}\tilde{w}_*} \ . \label{wf2phys22}
\end{equation}
We have to emphasize that the specific heat for type-II Dirac fermions at low temperatures is not
linear in $T$ because $v_{1*}$, $v_{2*}$, and $\tilde{w}_*$ are functions of $T$. This deviation from
the linear $T$ behavior suggests the effect of Coulomb interactions.

\section{Screening of the Coulomb potential}
\label{coul}

One may wonder whether or not the above RG flows we obtained are cut at low energies due to the
screening of the long range Coulomb interaction. To answer this question, we compute the vacuum
polarization $\Pi(Q)$ to the one-loop order, where the vacuum polarization is defined by the Dyson
equation
\begin{equation}
 D^{-1}(Q)=\mathcal{V}^{-1}_0(\bm{q})+\Pi(Q) \ . \label{wf2dd1}
\end{equation}
In Eq. (\ref{wf2dd1}), $D(Q)$ is the the full propagator of the $\phi$ field and
$\mathcal{V}_0(\bm{q})=\frac{g^2}{2|\bm{q}|}$ is the bare $\phi$ propagator as well as the Fourier
transform of the bare Coulomb potential $V_0(\bm{r})$. The Fourier transform $\mathcal{V}_s(\bm{q})$ of
the renormalized Coulomb potential is then given by $\mathcal{V}_s(\bm{q})=D(0,\bm{q})$.

\subsection{The screened Coulomb potential}

As we have shown, our RG equations for type-I Dirac fermions are identical to those by regularizing the
theory with a momentum cutoff. Hence, we may compute $\Pi(Q)$ in terms of dimensional
regularization\cite{DTSon}, yielding
\begin{equation}
 \Pi(Q)=\frac{N}{16v_1v_2} \! \sum_{\xi}\frac{\tilde{\bm{q}}^2}
 {\sqrt{(q_0+i\xi w\tilde{q}_1)^2+\tilde{\bm{q}}^2}} \ . \label{wf2dd11}
\end{equation}
When $w=0$, Eq. (\ref{wf2dd11}) reduces to the one for graphene\cite{DTSon}. On the other hand, for
type-II Dirac fermions, we have to employ our parametrization for momenta [Eq. (\ref{wf2e11})]. An
exact evaluation of $\Pi(Q)$ is difficult. Fortunately, to answer the question of screening, it
suffices to determine $\Pi(0,\bm{q})$, and we find that for $v_1|q_1|,v_2|q_2|\ll D$
\begin{equation}
 \Pi(0,\bm{q})=2\tilde{w}^2BD(0)\frac{(w^2+1)r^2q_1^2+q_2^2}{(w^2+1)^2r^2q_1^2-\tilde{w}^2q_2^2} \ ,
 \label{wf2dd12}
\end{equation}
where $D$ is the band width, $r=v_1/v_2$, $D(0)=\frac{ND}{4\pi^2v_1v_2\tilde{w}}$ is the DOS at the
Fermi level, and $B$ is a nonuniversal constant. The details of the calculations are left to appndix
\ref{pol}.

Equations (\ref{wf2dd11}) and (\ref{wf2dd12}) indicate that
\begin{eqnarray*}
 \lim_{q_1\rightarrow 0}\lim_{q_2\rightarrow 0}\Pi(0,\bm{q})\neq\lim_{q_2\rightarrow 0}
 \lim_{q_1\rightarrow 0}\Pi(0,\bm{q}) \ ,
\end{eqnarray*}
for type-II Dirac fermions and type-I Dirac fermions with $w\neq 0$. This implies that $\bm{q}=0$ is a
singular point of $\Pi(0,\bm{q})$ as well as $\mathcal{V}_s(\bm{q})$.

For type-II Dirac fermions, we remove this singularity by summing the leading divergent diagrams,
following the method employed in Ref. \onlinecite{RRBiswas}. Since this theory is renormalizable, this
can be achieved by replacing $w$ and $r$ in Eq. (\ref{wf2dd12}) by the energy dependent functions $w_l$
and $r_l$ and scaling $w_l$ and $r_l$ to the energy scale $\sqrt{\tilde{w}^2v_1^2q_1^2+v_2^2q_2^2}$.
Thus, the polarization function becomes
\begin{equation}
 \Pi(0,\bm{q})=2\tilde{w}_l^2BD(0)\frac{(w_l^2+1)r_l^2q_1^2+q_2^2}
 {(w_l^2+1)^2r_l^2q_1^2-\tilde{w}_l^2q_2^2} \ . \label{wf2dd13}
\end{equation}
As we have shown, $|w_l|$ is an increasing function of $l$, while $r_l$ is a decreasing function of $l$.
Moreover, the product $|w_l|r_l$ increases with increasing $l$. Hence, at low energies (corresponding to
small momenta), we may take $|w_l|\gg 1$ such that Eq. (\ref{wf2dd13}) can be approximated as
$\Pi(0,\bm{q})\approx 2BD(0)$. In terms of this expression, $\mathcal{V}_s(\bm{q})$ becomes
\begin{equation}
 \mathcal{V}_s(\bm{q})=\frac{g^2/2}{q+q_{TF}} \ , \label{wf2v1}
\end{equation}
which holds for $q=|\bm{q}|\ll q_{TF}$, where $q_{TF}=BD(0)g^2$ is the Thomas-Fermi wavenumber. When
$r\gg 1/q_{TF}$, the screened Coulomb potential $V_s(\bm{r})$ behaves like
\begin{equation}
 V_s(\bm{r})\approx\frac{e^2}{4\pi\epsilon q_{TF}^2r^3} \ , \label{wf2v11}
\end{equation}
instead of the bare one $V_0(\bm{r})\sim 1/r$. This indicates that the RG flow we have obtained holds
only when the energy scale is much larger than $v_0q_{TF}$, where we have taken $v_1=v_0=v_2$. On
account of the renormalization of the Coulomb interaction, $D(0)$ is smaller than the value for
non-interacting fermions, and thus we expect that our results for the compressibility and specific heat
hold for a large temperature range. Notice that the anisotropy of the screened Coulomb potential due to
tilting is removed at long distances.

\subsection{Coulomb impurity}

In terms of the above polarization function, we can also study the Coulomb impurity probelm for tilted
Dirac fermions. Consider an impurity of charge $Ze$ located at the origin, where the charge carried by
an electron is $-e$. The charge density induced by this impruity is given by
\begin{eqnarray*}
 \rho_{\mbox{ind}}(\bm{r})=\! \int \! \frac{d^2q}{(2\pi)^2}\tilde{\rho}_{\mbox{ind}}(\bm{q})
 e^{i\bm{q}\cdot\bm{r}} \ ,
\end{eqnarray*}
where
\begin{equation}
 \tilde{\rho}_{\mbox{ind}}(\bm{q})=-Ze\Pi(0,\bm{q})\mathcal{V}_s(\bm{q}) \ , \label{wf2ind1}
\end{equation}

For type-II Dirac fermions, we employ Eq. (\ref{wf2v1}) and get
\begin{equation}
 \tilde{\rho}_{\mbox{ind}}(\bm{q})=-\frac{q_{TF}}{q+q_{TF}}Ze \ , \label{wf2ind11}
\end{equation}
which holds only for $q\ll q_{TF}$. The total induced charge $Q_{\mbox{ind}}$ is then given by
\begin{eqnarray*}
 Q_{\mbox{ind}}=\! \int \!d^2r\rho_{\mbox{ind}}(\bm{r})=\tilde{\rho}_{\mbox{ind}}(0)=-Ze \ ,
\end{eqnarray*}
which implies the complete screening of the impurity charge. The presence of a nonvanishing screening
length $1/q_{TF}$ and the complete screening of the impurity charge rely on a nonzero DOS at the Fermi
energy. In contrast with the usual FL with a finite Fermi momentum, for type-II Dirac fermions, we have
to sum the leading divergent diagrams beyond the RPA approximation.

For type-I Dirac fermions, we obtain
\begin{equation}
 \tilde{\rho}_{\mbox{ind}}(\bm{q})=-\frac{\lambda Zeq}{\sqrt{(1-w^2)q_1^2+q_2^2}+\lambda q} \ ,
 \label{wf2ind12}
\end{equation}
within the RPA approximation, where for simplicity we have set $v_1=v=v_2$ and $\lambda=Ng^2/(16v)$.
When $w\neq 0$, we find that
\begin{eqnarray*}
 \lim_{q_1\rightarrow 0}\lim_{q_2\rightarrow 0}\tilde{\rho}_{\mbox{ind}}(\bm{q})\neq
 \lim_{q_2\rightarrow 0}\lim_{q_1\rightarrow 0}\tilde{\rho}_{\mbox{ind}}(\bm{q}) \ .
\end{eqnarray*}
That is, $\bm{q}=0$ is a singular point of $\tilde{\rho}_{\mbox{ind}}(\bm{q})$ when $w\neq 0$. This
singularity can also be removed by summing the leading logarithmically divergent diagrams. This can be
achieved by replacing $\lambda$ and $w$ by $\lambda(p)$ and $w(p)$ and scaling $\lambda(p)$ and $w(p)$
to the momentum scale $p=q$. As a result, we get
\begin{equation}
 \tilde{\rho}_{\mbox{ind}}(\bm{q})=-\frac{q\lambda(q) Ze}{\sqrt{[1-w^2(q)]q_1^2+q_2^2}+q\lambda(q)} \ .
 \label{wf2ind13}
\end{equation}
Since $w(0)=0$ and $\lambda(0)=0$, we conclude that $Q_{\mbox{ind}}=\tilde{\rho}_{\mbox{ind}}(0)=0$. In
other words, the presence of an ion only leads to charge redistribution: a fraction of $Z$ charge is
pushed from short distances (of order of lattice spacing) to longer distances, but none of the charge
goes to infinity. This situation is similar to graphene\cite{RRBiswas}.

\section{Conclusions and discussions}
\label{condis}

In the present work, we study the effects of Coulomb interactions on the tilted Dirac fermions in $2$D
with the help of RG. For type-I Dirac fermions, our method leads to identical results to previous
studies. For type-II Dirac fermions, however, we find that the Coulomb interaction helps to stabilize
the type-II Dirac semimetals, in contrast with previous works. Since our approach is perturbative in
nature, the results hold only in the weak-coupling regime. Although we focus on tilted Dirac fermions
in $2$D, the extension of our method to tilted Weyl fermions in $3$D is straightforward. In fact, the
parametrization of momenta is similar to Eqs. (\ref{wf2e1}) or (\ref{wf2e11}), except the introduction
of an additional ``angle" variable.

With the help of the RG equations, we can study thermodynamics of tilted Dirac fermions at low
temperatures and/or small densities. In particular, we calculate the temperature or density dependence
of the isothermal compressibility and specific heat for both types of Dirac fermions.

To answer the question of screening of the Coulomb potential, we compute the vacuum polarization at
zero frequency to the one-loop order for both types of Dirac fermions. For type-II Dirac fermions, we
obtain a finite screening length, leading to total screening of a charged impurity, similar to the
usual FL. Such a conclusion is obtained within the RPA approximation for the usual FL. In contrast, for
type-II Dirac fermions, we have to go beyond the RPA approximation by summing the leading divergent
diagrams. Since the screening length is inversely proportional to the DOS at the Fermi level and the
latter is suppressed by Coulomb interactions, the screening length is larger than the one from the
estimate based on non-interacting fermions. This implies that our results on the temperature or density
dependence of the isothermal compressibility and specific heat for type-II Dirac fermions hold for a
large temperature range. As for type-I Dirac fermions, the results are similar to those for graphene
after summing the leading logarithmically divergent diagrams. The only effect of $w\neq 0$ is that the
induced charge density becomes anisotropic due to the breaking of rotational symmetry.

The major difference between our work and the previous ones is that we adopt the regularization scheme
such that the RG transformation is to scale to the Fermi surface for both types of Dirac fermions.
This is important, in particular, for type-II Dirac fermions since its Fermi surface is open. This
point is also noticed by a recent work on tilted Weyl fermions in $3$D\cite{FDetassis}.

Another possible way to study this problem properly is to work directly on the low-energy effective
theory for the fermionic excitations near the Fermi surface, as was done in Ref. \onlinecite{ZMHuang}.
However, the conclusions in that work are distinct from ours. In our opinion, this distinction may
arise from the following reason. The dimensional regularization employed there is a
Lorentz/rotational-invariant regularization scheme which treats the frequency $p_0$,
the momenta parallel ($p_{\parallel}$) and perpendicular ($p_{\perp}$) to the Fermi lines on equal
footing. However, the role played by $p_{\parallel}$ is to parametrize the location of points on the
Fermi lines, as what our parameter $\theta$ does. Fermionic excitations with different values of
$p_{\parallel}$, but the same value of $p_{\perp}$, must have the same energy. Hence, the scaling of
$p_{\parallel}$ under the RG transformations can not be the same as the one of $p_{\perp}$ since
$p_{\perp}$ scales to zero while $p_{\parallel}$ does not. Accordingly, the use of dimensional
regularization here is problematic.

An interesting question to ask is whether or how does the inclusion of transverse gauge
fluctuations\cite{HIsobe2} by using our approach modify the above results. For the type-I case, it was
shown that the velocities of the Dirac fermions are renormalized toward the speed of light and the
tilting parameter flows to zero. Therefore, the Lorentz symmetry is restored at low energies.
The situation is more subtle for type-II Dirac fermions since when the tilting parameter $\abs{w}>1$, 
the Fermi surface is not point-like and type-II Dirac fermions is separated from type-I Dirac fermions 
by a quantum phase transition point. Due to this intrinsic breaking of the Lorentz/rotational symmetry, 
there is a priori no reason to expect that the same restoration of Lorentz symmetry will occur in the 
type-II case. Moreover, previous studies on the FL with a closed Fermi surface in $2$D, interacting 
with transverse U($1$) gauge fields, suggests that the low-energy physics is controlled by an 
interacting fixed point\cite{Polch,Nayak,Hermele,SSLee,DFMross}. Since the fate of type-II Dirac 
fermions interacting with transverse U($1$) gauge fields is a dynamical issue, we leave it for future 
study.

\acknowledgments

The works of Y.L. Lee and Y.-W. Lee are supported by the Ministry of Science and Technology, Taiwan,
under the grant number MOST 106-2112-M-018-003 and MOST 106-2112-M-029-002, respectively.

\appendix

\section{The self-energy of tilted Dirac fermions}
\label{self}

Here we present the derivation of the one-loop RG equations. As we have discussed, it suffices to
compute the self-energy $\Sigma_{\xi}(K)$ of fermions. To proceed, we need the free propagator of
fermions:
\begin{eqnarray*}
 & & \langle\mathcal{T}_{\tau}\{\psi_{\xi\alpha}(X_1)\psi^{\dagger}_{\xi^{\prime}\alpha^{\prime}}(X_2)
     \}\rangle \\
 & & =\delta_{\xi\xi^{\prime}}\delta_{\alpha\alpha^{\prime}} \! \int_P
	 e^{-ip_0(\tau_1-\tau_2)+i\bm{p}\cdot(\bm{r}_1-\bm{r}_2)}G_{\xi 0}(P) \ ,
\end{eqnarray*}
and the free propagator of the bosonic field $\phi$:
\begin{eqnarray*}
 \langle\mathcal{T}_{\tau}\{\phi(X_1)\phi(X_2)\}\rangle =\! \int_P
 e^{-ip_0(\tau_1-\tau_2)+i\bm{p}\cdot(\bm{r}_1-\bm{r}_2)}D_0(\bm{p}) \ ,
\end{eqnarray*}
where $D_0(\bm{p})=\mathcal{V}_0(\bm{p})=\frac{g^2}{2|\bm{p}|}$ and
\begin{eqnarray*}
 G_{\xi 0}(P)=\frac{ip_0-\xi wv_1p_1+\xi v_1p_1\sigma_1+v_2p_2\sigma_2}
 {(p_0+i\xi wv_1p_1)^2+v_1^2p^2_1+v_2^2p^2_2} \ .
\end{eqnarray*}

By integrating out the fast modes, the following term
\begin{eqnarray*}
 \sum_{\xi,\alpha} \! \int_K\tilde{\psi}^{\dagger}_{\xi\alpha}(K)\Sigma_{\xi}(K)\tilde{\psi}_{\xi\alpha}
 (K) \ ,
\end{eqnarray*}
will be generated in the action $S$. To the one-loop order, we have
\begin{eqnarray*}
 & & \! \! \! \Sigma_{\xi}(K)=(-1)\cdot\frac{1}{2!}\cdot (-i)^2\cdot 2 \! \! \int_PG_{\xi 0}(P)D_0
	 (\bm{k}-\bm{p}) \\
 & & \! \! \! =\frac{g^2}{2} \! \!\int_P \! \frac{ip_0-\xi wv_1p_1+\xi v_1p_1\sigma_1+v_2p_2\sigma_2}
	{|\bm{k}-\bm{p}|[(p_0+i\xi wv_1p_1)^2+v_1^2p^2_1+v_2^2p^2_2]} \\
 & & \! \! \! =\Sigma_{\xi 0}(\bm{k})\sigma_0+\Sigma_{\xi 1}(\bm{k})\sigma_1+\Sigma_{\xi 2}(\bm{k})
	 \sigma_2 \ ,
\end{eqnarray*}
where $\sigma_0$ is the $2\times 2$ unit matrix,
\begin{eqnarray*}
 \Sigma_{\xi 0}(\bm{k}) &=& \frac{ig^2}{2v_1v_2} \! \int_{\mathcal{D}} \! \frac{d^2\tilde{p}}{(2\pi)^2}
 \frac{1}{\sqrt{\sum_{a=1,2}(k_a-\tilde{p}_a/v_a)^2}} \\
 & & \times \! \int^{+\infty}_{-\infty} \! \frac{dp_0}{2\pi}\frac{p_0+i\xi w\tilde{p}_1}
 {(p_0+i\xi w\tilde{p}_1)^2+\tilde{\bm{p}}^2} \\
 &=& \frac{g^2}{8v_1v_2} \! \int_{\mathcal{D}} \! \frac{d^2\tilde{p}}{(2\pi)^2}
 \frac{\mbox{sgn}[\epsilon_+(\bm{p})]+\mbox{sgn}[\epsilon_-(\bm{p})]}
 {\sqrt{\sum_{a=1,2}(k_a-\tilde{p}_a/v_a)^2}} \ ,
\end{eqnarray*}
and
\begin{eqnarray*}
 \Sigma_{\xi 1}(\bm{k}) &=& \frac{\xi g^2}{2v_1v_2} \! \int_{\mathcal{D}} \! \frac{d^2\tilde{p}}
 {(2\pi)^2}\frac{\tilde{p}_1}{\sqrt{\sum_{a=1,2}(k_a-\tilde{p}_a/v_a)^2}} \\
 & & \times \! \int^{+\infty}_{-\infty} \! \frac{dp_0}{2\pi}\frac{1}
 {(p_0+i\xi w\tilde{p}_1)^2+\tilde{\bm{p}}^2} \\
 &=& \frac{\xi g^2}{8v_1v_2} \! \! \int_{\mathcal{D}} \! \frac{d^2\tilde{p}}{(2\pi)^2}
 \frac{\tilde{p}_1[\mbox{sgn}(\epsilon_+(\bm{p}))-\mbox{sgn}(\epsilon_-(\bm{p}))]}
 {\tilde{p}\sqrt{\sum_{a=1,2}(k_a-\tilde{p}_a/v_a)^2}} \ , \\
 \Sigma_{\xi 2}(\bm{k}) &=& \frac{g^2}{2v_1v_2} \! \int_{\mathcal{D}} \! \frac{d^2\tilde{p}}{(2\pi)^2}
 \frac{\tilde{p}_2}{\sqrt{\sum_{a=1,2}(k_a-\tilde{p}_a/v_a)^2}} \\
 & & \times \! \int^{+\infty}_{-\infty} \! \frac{dp_0}{2\pi}\frac{1}
 {(p_0+i\xi w\tilde{p}_1)^2+\tilde{\bm{p}}^2} \\
 &=& \frac{g^2}{8v_1v_2} \! \! \int_{\mathcal{D}} \! \frac{d^2\tilde{p}}{(2\pi)^2}
 \frac{\tilde{p}_2[\mbox{sgn}(\epsilon_+(\bm{p}))-\mbox{sgn}(\epsilon_-(\bm{p}))]}
 {\tilde{p}\sqrt{\sum_{a=1,2}(k_a-\tilde{p}_a/v_a)^2}} \ .
\end{eqnarray*}
In the above, $\mathcal{D}$ is the energy shell in the range $\Lambda/s<|E|<\Lambda$.

\subsection{Type-I Dirac fermions}

We first consider type-I Dirac fermions. In this case, $\epsilon_+(\bm{p})>0$ and $\epsilon_-(\bm{p})<0$
as long as $E\neq 0$. Thus, the above expressions reduce to
\begin{eqnarray*}
 \Sigma_{\xi 1}(\bm{k}) \! \! &=& \! \! \frac{\xi g^2}{4v_1v_2} \! \! \int_{\mathcal{D}} \!
 \frac{d^2\tilde{p}}{(2\pi)^2}\frac{\tilde{p}_1}{\tilde{p}\sqrt{\sum_{a=1,2}(k_a-\tilde{p}_a/v_a)^2}}
 \ , \\
 \Sigma_{\xi 2}(\bm{k}) \! \! &=& \! \! \frac{g^2}{4v_1v_2} \! \! \int_{\mathcal{D}} \!
 \frac{d^2\tilde{p}}{(2\pi)^2}\frac{\tilde{p}_2}{\tilde{p}\sqrt{\sum_{a=1,2}(k_a-\tilde{p}_a/v_a)^2}}
 \ , \\
 \Sigma_{\xi 0}(\bm{k}) \! \! &=& \! \! 0 \ .
\end{eqnarray*}
To proceed, we expand $\Sigma_{\xi a}(\bm{k})$ with $a=1,2$ to the linear orders in $\bm{k}$:
\begin{eqnarray*}
 \Sigma_{\xi a}(\bm{k})=\Sigma_{\xi a}^{(0)}+\! \sum_{i=1,2}\Sigma_{\xi a}^{(i)}k_i+O(k_i^2) \ ,
\end{eqnarray*}
where
\begin{eqnarray*}
 \Sigma_{\xi 1}^{(0)} &=& \frac{\xi g^2}{4v^2_1v_2} \! \! \int_{\mathcal{D}} \! \frac{d^2\tilde{p}}
 {(2\pi)^2}\frac{\tilde{p}_1}{\tilde{p}\sqrt{(\tilde{p}_1/v_1)^2+(\tilde{p}_2/v_2)^2}} \\
 &=& 0 \ , \\
 \Sigma_{\xi 2}^{(0)} &=& \frac{\xi g^2}{4v^2_1v_2} \! \! \int_{\mathcal{D}} \! \frac{d^2\tilde{p}}
 {(2\pi)^2}\frac{\tilde{p}_2}{\tilde{p}\sqrt{(\tilde{p}_1/v_1)^2+(\tilde{p}_2/v_2)^2}} \\
 &=& 0 \ , \\
 \Sigma_{\xi 1}^{(1)} &=& \frac{\xi g^2}{4v^2_1v_2} \! \! \int_{\mathcal{D}} \! \frac{d^2\tilde{p}}
 {(2\pi)^2}\frac{\tilde{p}_1^2}{\tilde{p}[(\tilde{p}_1/v_1)^2+(\tilde{p}_2/v_2)^2]^{3/2}} \ , \\
 \Sigma_{\xi 1}^{(2)} &=& \frac{\xi g^2}{4v_1v_2^2} \! \! \int_{\mathcal{D}} \! \frac{d^2\tilde{p}}
 {(2\pi)^2}\frac{\tilde{p}_1\tilde{p}_2}{\tilde{p}[(\tilde{p}_1/v_1)^2 \! +(\tilde{p}_2/v_2)^2]^{3/2}}
 \\
 &=& 0 \ , \\
 \Sigma_{\xi 2}^{(1)} &=& \frac{g^2}{4v^2_1v_2} \! \! \int_{\mathcal{D}} \! \frac{d^2\tilde{p}}
 {(2\pi)^2}\frac{\tilde{p}_1\tilde{p}_2}{\tilde{p}[(\tilde{p}_1/v_1)^2+(\tilde{p}_2/v_2)^2]^{3/2}} \\
 &=& 0 \ , \\
 \Sigma_{\xi 2}^{(2)} &=& \frac{g^2}{4v_1v_2^2} \! \! \int_{\mathcal{D}} \! \frac{d^2\tilde{p}}
 {(2\pi)^2}\frac{\tilde{p}_2^2}{\tilde{p}[(\tilde{p}_1/v_1)^2+(\tilde{p}_2/v_2)^2]^{3/2}} \ .
\end{eqnarray*}
$\Sigma_{\xi 1}^{(2)}=0=\Sigma_{\xi 2}^{(0)}=\Sigma_{\xi 2}^{(1)}$ because $\mathcal{D}$ is symmetric
under the reflection $\tilde{p}_2\rightarrow-\tilde{p}_2$, while the integrands are odd functions of
$\tilde{p}_2$. On the other hand, $\Sigma_{\xi 1}^{(0)}=0$ by an explicit calculation. The
renormalized velocities $v_1^{\prime}$ and $v_2^{\prime}$ are then given by
\begin{eqnarray*}
 v_1^{\prime}=v_1+\xi\Sigma_{\xi 1}^{(1)} \ , ~~v_2^{\prime}=v_2+\Sigma_{\xi 2}^{(2)} \ .
\end{eqnarray*}

For simplicity, we consider $v_1=v_2$, and we have
\begin{eqnarray*}
 \Sigma_{\xi 1}^{(1)} &=& \frac{\xi\sqrt{1-w^2}g^2}{32\pi^2} \! \! \int^{\Lambda}_{\Lambda/s}
 \frac{dE}{E} \! \! \int^{2\pi}_0 \! \frac{(w-\xi\cos{\theta})^2d\theta}{(1-\xi w\cos{\theta})^3} \\
 & & +\frac{\xi\sqrt{1-w^2}g^2}{32\pi^2} \! \! \int_{-\Lambda}^{-\Lambda/s} \! \frac{dE}{|E|} \! \!
 \int^{2\pi}_0 \! \frac{(w+\xi\cos{\theta})^2d\theta}{(1+\xi w\cos{\theta})^3} \\
 &=& \frac{\xi g^2l}{16\pi} \ ,
\end{eqnarray*}
and
\begin{eqnarray*}
 \Sigma_{\xi 2}^{(2)} &=& \frac{(1-w^2)^{3/2}g^2}{32\pi^2} \! \! \int^{\Lambda}_{\Lambda/s}
 \frac{dE}{E} \! \! \int^{2\pi}_0 \! \frac{\sin^2{\theta}d\theta}{(1-\xi w\cos{\theta})^3} \\
 & & +\frac{(1-w^2)^{3/2}g^2}{32\pi^2} \! \! \int_{-\Lambda}^{-\Lambda/s} \! \frac{dE}{|E|} \! \!
 \int^{2\pi}_0 \! \frac{\sin^2{\theta}d\theta}{(1+\xi w\cos{\theta})^3} \\
 &=& \frac{g^2l}{16\pi} \ .
\end{eqnarray*}
As a result, the renormalized velocity to the one-loop order is given by
\begin{eqnarray*}
 v^{\prime}=v+\frac{g^2l}{16\pi} \ ,
\end{eqnarray*}
which gives the RG equation for $v$ in Eq. (\ref{wf2rge2}). On the other hand, $wv_1$ is not
renormalized to the one-loop order because $\Sigma_{\xi 0}=0$. Thus, we get
\begin{eqnarray*}
 (wv_1)^{\prime}=wv_1 \ ,
\end{eqnarray*}
which leads to the RG equation for $wv$ in Eq. (\ref{wf2rge2}).

\subsection{Type-II Dirac fermions}

Next, we consider type-II Dirac fermions. In terms of the parametrization (\ref{wf2e11}), we find that
\begin{eqnarray*}
 \mbox{sgn}(\epsilon_+) &=& \mbox{sgn}(E)+2\Theta(\pm\eta_w\xi)\Theta(-E)\Theta(|\theta|-\ln{|w|}) \ ,
 \\
 \mbox{sgn}(\epsilon_-) &=& \mbox{sgn}(E)-2\Theta(\mp\eta_w\xi)\Theta(E)\Theta(|\theta|-\ln{|w|}) \ ,
\end{eqnarray*}
where the upper and the lower signs correspond to the right and the left branches of the hyperbola,
respectively. To compute the renormalization of various parameters in the action, we expand
$\Sigma_{\xi\mu}(\bm{k})$ with $\mu=0,1,2$ to the linear orders in $\bm{k}$:
\begin{eqnarray*}
 \Sigma_{\xi\mu}(\bm{k})=\Sigma_{\xi\mu}^{(0)}+\! \sum_{i=1,2}\Sigma_{\xi\mu}^{(i)}k_i+O(k_i^2) \ ,
\end{eqnarray*}
Among these terms, $\Sigma_{\xi 0}^{(2)}=0=\Sigma_{\xi 1}^{(2)}=\Sigma_{\xi 2}^{(0)}=\Sigma_{\xi 2}^{(1)}$
because the integration domain is symmetric under the reflection $\tilde{p}_2\rightarrow -\tilde{p}_2$,
while the integrand is an odd function of $\tilde{p}_2$. On the other hand,
$\Sigma_{\xi 0}^{(0)}=0=\Sigma_{\xi 1}^{(0)}$ by explicit claculations. Thus, the nonvanishing terms are
\begin{eqnarray*}
 \Sigma_{\xi 0}^{(1)} &=& \frac{g^2}{32\pi^2v_1^2v_2} \! \int_{\mathcal{D}} \! d^2\tilde{p}
 \frac{\tilde{p}_1[\mbox{sgn}(\epsilon_+)+\mbox{sgn}(\epsilon_-)]}
 {[(\tilde{p}_1/v_1)^2+(\tilde{p}_2/v_2)^2]^{3/2}} \ , \\
 \Sigma^{(1)}_{\xi 1} &=& \frac{\xi g^2}{32\pi^2v_1^2v_2} \! \int_{\mathcal{D}} \! d^2\tilde{p}
 \frac{\tilde{p}_1^2[\mbox{sgn}(\epsilon_+)-\mbox{sgn}(\epsilon_-)]}
 {\tilde{p}[(\tilde{p}_1/v_1)^2+(\tilde{p}_2/v_2)^2]^{3/2}} \ , \\
 \Sigma^{(2)}_{\xi 2} &=& \frac{g^2}{32\pi^2v_1v_2^2} \! \int_{\mathcal{D}} \! d^2\tilde{p}
 \frac{\tilde{p}_2^2[\mbox{sgn}(\epsilon_+)-\mbox{sgn}(\epsilon_-)]}
 {\tilde{p}[(\tilde{p}_1/v_1)^2+(\tilde{p}_2/v_2)^2]^{3/2}} \ .
\end{eqnarray*}
The renormalized values of $wv_1$, $v_1$, and $v_2$ to the one-loop order are then given by
\begin{eqnarray*}
 (wv_1)^{\prime} &=& wv_1+\xi\Sigma_{\xi 0}^{(1)} \ , \\
 v_1^{\prime} &=& v_1+\xi\Sigma_{\xi 1}^{(1)} \ , \\
 v_2^{\prime} &=& v_2+\Sigma_{\xi 2}^{(2)} \ .
\end{eqnarray*}

In terms of Eqs. (\ref{wf2e11}) and (\ref{wf2e13}), we find that
\begin{eqnarray*}
 (wv_1)^{\prime} &=& wv_1+\frac{\eta_w\tilde{w}g^2rl}{8\pi^2}(J_1-J_2) \ , \\
 v_1^{\prime} &=& v_1+\frac{\tilde{w}g^2rl}{8\pi^2}M_1 \ , \\
 v_2^{\prime} &=& v_2+\frac{\tilde{w}^3g^2r^2l}{8\pi^2}M_2 \ ,
\end{eqnarray*}
where
\begin{eqnarray*}
 J_1 &=& \! \int^{+\infty}_0 \! d\theta \! \left\{\frac{(|w|\cosh{\theta}+1)(|w|+\cosh{\theta})}
 {[(|w|+\cosh{\theta})^2+r^2\tilde{w}^2\sinh^2{\theta}]^{3/2}}\right. \\
 & & \! \left. +\frac{(|w|\cosh{\theta}-1)(|w|-\cosh{\theta})}
 {[(|w|-\cosh{\theta})^2+r^2\tilde{w}^2\sinh^2{\theta}]^{3/2}}\right\} , \\
 J_2 &=& \! \int^{+\infty}_{\ln{|w|}} \! d\theta\frac{(|w|\cosh{\theta}-1)(|w|-\cosh{\theta})}
 {[(|w|-\cosh{\theta})^2+r^2\tilde{w}^2\sinh^2{\theta}]^{3/2}} \ .
\end{eqnarray*}
while $M_1$ and $M_2$ are defined in Sec. \ref{rge}. From the above equations, we obtain Eqs.
(\ref{wf2rge21}) -- (\ref{wf2rge23}) with the function $\mathcal{N}=J_1-J_2-|w|M_1$.

\section{The vacuum polarization}
\label{pol}

Here we compute the vacuum polarization to the one-loop order, which is given by
\begin{widetext}
\begin{eqnarray*}
 & & \Pi(Q)=(-2)\cdot\frac{1}{2!}\cdot (-i)^2\cdot (-1)\cdot N \! \sum_{\xi} \! \int_P
	 \mbox{tr}[G_{\xi 0}(P)G_{\xi 0}(P+Q)] \\
 & & =\frac{2N}{v_1v_2} \! \sum_{\xi} \! \int \! \frac{d^2\tilde{p}}{(2\pi)^2} \!
     \int^{+\infty}_{-\infty} \! \frac{dp_0}{2\pi}
     \frac{(p_0+i\xi w\tilde{p}_1)[p_0+q_0+i\xi w(\tilde{p}_1+\tilde{q}_1)]
	 -\tilde{\bm{p}}\cdot(\tilde{\bm{p}}+\tilde{\bm{q}})}{[p_0+i\epsilon_+(\bm{p})]
	 [p_0+i\epsilon_-(\bm{p})][p_0+q_0+i\epsilon_+(\bm{p}+\bm{q})]
	 [p_0+q_0+i\epsilon_-(\bm{p}+\bm{q})]} \ ,
\end{eqnarray*}
where $\tilde{\bm{q}}=(v_1q_1,v_2q_2)$.

For type-I Dirac fermions, we employ the Feynmann parametrization
\begin{eqnarray*}
	\frac{1}{ab}=\! \int^1_0 \! \frac{dx}{[ax+b(1-x)]^2} \ ,
\end{eqnarray*}
and write $\Pi(Q)$ as
\begin{eqnarray*}
 \Pi(Q) &=& \frac{2N}{v_1v_2} \! \sum_{\xi} \! \int^1_0 \!dx \! \int^{+\infty}_{-\infty} \! \frac{d^3P}
 {(2\pi)^3}\frac{(p_0+i\xi w\tilde{p}_1)[p_0+q_0+i\xi w(\tilde{p}_1+\tilde{q}_1)]
 -\tilde{\bm{p}}\cdot(\tilde{\bm{p}}+\tilde{\bm{q}})}
 {\{[(p_0+i\xi w\tilde{p}_1)^2+\tilde{\bm{p}}^2](1-x)+x[(p_0+q_0+i\xi w(\tilde{p}_1+\tilde{q}_1))^2
 +(\tilde{\bm{p}}+\tilde{\bm{q}})^2]\}^2} \\
 &=& \frac{2N}{v_1v_2} \! \sum_{\xi} \! \int^1_0 \!dx \! \int^{+\infty}_{-\infty} \! \frac{d^3P}
 {(2\pi)^3}\frac{(p_0+i\xi w\tilde{p}_1)[p_0+q_0+i\xi w(\tilde{p}_1+\tilde{q}_1)]
 -\tilde{\bm{p}}\cdot(\tilde{\bm{p}}+\tilde{\bm{q}})}
 {\{[p_0+i\xi w\tilde{p}_1+x(q_0+i\xi w\tilde{q}_1)]^2+(\tilde{\bm{p}}+x\tilde{\bm{q}})^2+x(1-x)
 [(q_0+i\xi w\tilde{q}_1)^2+\tilde{\bm{q}}^2]\}^2} \\
 &=& \frac{2N}{v_1v_2} \! \sum_{\xi} \! \int^1_0 \!dx \! \int^{+\infty}_{-\infty} \! \frac{d^3P}
 {(2\pi)^3}\frac{1}
 {\{(p_0+i\xi w\tilde{p}_1)^2+\tilde{\bm{p}}^2+x(1-x)[(q_0+i\xi w\tilde{q}_1)^2+\tilde{\bm{q}}^2]\}^2}
 \\
 & & \times\left\{(p_0+i\xi w\tilde{p}_1)^2-\tilde{\bm{p}}^2
 +(1-2x)[(p_0+i\xi w\tilde{p}_1)(q_0+i\xi w\tilde{q}_1)-\tilde{\bm{p}}\cdot\tilde{\bm{q}}
 -x(1-x)[(q_0+i\xi w\tilde{q}_1))^2-\tilde{\bm{q}}^2]\right\} \\
 &=& \frac{2N}{v_1v_2} \! \sum_{\xi} \! \int^1_0 \!dx \! \int^{+\infty}_{-\infty} \! \frac{d^3P}{(2\pi)^3}
 \frac{p_0^2-\tilde{\bm{p}}^2-x(1-x)[(q_0+i\xi w\tilde{q}_1))^2-\tilde{\bm{q}}^2]}
 {\{P^2+x(1-x)[(q_0+i\xi w\tilde{q}_1)^2+\tilde{\bm{q}}^2]\}^2} \\
 &=& -\frac{2N}{v_1v_2} \! \sum_{\xi} \! \int^1_0 \!dx \! \int^{+\infty}_{-\infty} \! \frac{d^3P}
 {(2\pi)^3}\frac{P^2/3+x(1-x)[(q_0+i\xi w\tilde{q}_1))^2-\tilde{\bm{q}}^2]}
 {\{P^2+x(1-x)[(q_0+i\xi w\tilde{q}_1)^2+\tilde{\bm{q}}^2]\}^2} \ ,
\end{eqnarray*}
where $d^3P=dp_0d^2\tilde{p}$ and $P^2=p_0^2+\tilde{\bm{p}}^2$. We regularize the momentum integral in
terms of the dimensional regularization, yielding
\begin{eqnarray*}
 \Pi(Q)=\frac{N}{2\pi v_1v_2} \! \sum_{\xi}\frac{\tilde{\bm{q}}^2}
 {\sqrt{(q_0+i\xi w\tilde{q}_1)^2+\tilde{\bm{q}}^2}} \! \int^1_0 \!dx\sqrt{x(1-x)} \ .
\end{eqnarray*}
Performing the $x$ integral, we get Eq. (\ref{wf2dd11}).

For type-II Dirac fermions, we first perform the $p_0$ integral, yielding
\begin{eqnarray*}
 \Pi(Q) &=& -\frac{2N}{v_1v_2} \! \sum_{\xi} \! \int \! \frac{d^2\tilde{p}}{(2\pi)^2}
 \frac{\mbox{sgn}[\epsilon_-(\bm{p})][\tilde{p}(\xi w\tilde{q}_1-iq_0)+\tilde{\bm{p}}^2
 +\tilde{\bm{p}}\cdot(\tilde{\bm{p}}+\tilde{\bm{q}})]}{\tilde{p}
 [\epsilon_+(\bm{p}+\bm{q})-\epsilon_-(\bm{p})-iq_0][\epsilon_-(\bm{p}+\bm{q})-\epsilon_-(\bm{p})-iq_0]}
 \\
 & & -\frac{2N}{v_1v_2} \! \sum_{\xi} \! \int \! \frac{d^2\tilde{p}}{(2\pi)^2}
 \frac{\mbox{sgn}[\epsilon_+(\bm{p})][\tilde{p}(\xi w\tilde{q}_1-iq_0)-\tilde{\bm{p}}^2
 -\tilde{\bm{p}}\cdot(\tilde{\bm{p}}+\tilde{\bm{q}})]}{\tilde{p}
 [\epsilon_+(\bm{p}+\bm{q})-\epsilon_+(\bm{p})-iq_0][\epsilon_-(\bm{p}+\bm{q})-\epsilon_+(\bm{p})-iq_0]}
 \ .
\end{eqnarray*}
It is straightforward to verify that $\Pi(q_0,0)=0$. On the other hand, $\Pi(0,\bm{q})$ can be written
as
\begin{eqnarray*}
 \Pi(0,\bm{q}) &=& -\frac{2N}{v_1v_2} \! \sum_{\xi} \! \int \! \frac{d^2\tilde{p}}{(2\pi)^2}
 \frac{\mbox{sgn}[\epsilon_-(\bm{p})][\xi w\tilde{q}_1\tilde{p}+\tilde{\bm{p}}^2
 +\tilde{\bm{p}}\cdot(\tilde{\bm{p}}+\tilde{\bm{q}})]}
 {\tilde{p}(\xi w\tilde{q}_1+|\tilde{\bm{p}}+\tilde{\bm{q}}|+\tilde{p})
 [\xi w\tilde{q}_1-(|\tilde{\bm{p}}+\tilde{\bm{q}}|-\tilde{p})]} \\
 & & -\frac{2N}{v_1v_2} \! \sum_{\xi} \! \int \! \frac{d^2\tilde{p}}{(2\pi)^2}
 \frac{\mbox{sgn}[\epsilon_+(\bm{p})][\xi w\tilde{q}_1\tilde{p}-\tilde{\bm{p}}^2
 -\tilde{\bm{p}}\cdot(\tilde{\bm{p}}+\tilde{\bm{q}})]}
 {\tilde{p}[\xi w\tilde{q}_1+(|\tilde{\bm{p}}+\tilde{\bm{q}}|-\tilde{p})]
 [\xi w\tilde{q}_1-(|\tilde{\bm{p}}+\tilde{\bm{q}}|+\tilde{p})]} \\
 &=& -\frac{N}{8\pi^2v_1v_2\tilde{w}}[F_1(\bm{q})-F_2(\bm{q})]-\frac{N}{4\pi^2v_1v_2\tilde{w}^3}
 [F_3(\bm{q})+F_4(\bm{q})] \ .
\end{eqnarray*}
where
\begin{eqnarray*}
 F_1(\bm{q}) &=& \! \sum_{\xi} \! \int^{\Lambda}_{-\Lambda} \! dE \! \int^{+\infty}_{-\infty} \! d\theta
 \frac{[|w|(\xi w\cosh{\theta}+\eta_E)+(\eta_E\xi w+\cosh{\theta})]\tilde{q}_1
 +\tilde{w}\tilde{q}_2\sinh{\theta}}{[|w|(\xi w\cosh{\theta}+\eta_E)-(\eta_E\xi w+\cosh{\theta})]
 \tilde{q}_1-\tilde{w}\tilde{q}_2\sinh{\theta}+\frac{\tilde{w}^2}{2|E|}
 (\tilde{w}^2\tilde{q}_1^2-\tilde{q}_2^2)} \\
 & & \times [\mbox{sgn}(E)-2\Theta(-\eta_w\xi)\Theta(E)\Theta(|\theta|-\ln{|w|})] \\
 & & +\! \sum_{\xi} \! \int^{\Lambda}_{-\Lambda} \! dE \! \int^{+\infty}_{-\infty} \! d\theta
 \frac{[|w|(\xi w\cosh{\theta}-\eta_E)+(\eta_E\xi w-\cosh{\theta})]\tilde{q}_1
 +\tilde{w}\tilde{q}_2\sinh{\theta}}{[|w|(\xi w\cosh{\theta}-\eta_E)-(\eta_E\xi w-\cosh{\theta})]
 \tilde{q}_1-\tilde{w}\tilde{q}_2\sinh{\theta}+\frac{\tilde{w}^2}{2|E|}
 (\tilde{w}^2\tilde{q}_1^2-\tilde{q}_2^2)} \\
 & & \times [\mbox{sgn}(E)-2\Theta(\eta_w\xi)\Theta(E)\Theta(|\theta|-\ln{|w|})] \ , \\
 F_2(\bm{q}) &=& \! \sum_{\xi} \! \int^{\Lambda}_{-\Lambda} \! dE \! \int^{+\infty}_{-\infty} \! d\theta
 \frac{[|w|(\xi w\cosh{\theta}+\eta_E)-(\eta_E\xi w+\cosh{\theta})]\tilde{q}_1
 -\tilde{w}\tilde{q}_2\sinh{\theta}}{[|w|(\xi w\cosh{\theta}+\eta_E)+(\eta_E\xi w+\cosh{\theta})]
 \tilde{q}_1+\tilde{w}\tilde{q}_2\sinh{\theta}-\frac{\tilde{w}^2}{2|E|}
 (\tilde{w}^2\tilde{q}_1^2-\tilde{q}_2^2)} \\
 & & \times [\mbox{sgn}(E)+2\Theta(\eta_w\xi)\Theta(-E)\Theta(|\theta|-\ln{|w|})] \\
 & & +\! \sum_{\xi} \! \int^{\Lambda}_{-\Lambda} \! dE \! \int^{+\infty}_{-\infty} \! d\theta
 \frac{[|w|(\xi w\cosh{\theta}-\eta_E)-(\eta_E\xi w-\cosh{\theta})]\tilde{q}_1
 -\tilde{w}\tilde{q}_2\sinh{\theta}}{[|w|(\xi w\cosh{\theta}-\eta_E)+(\eta_E\xi w-\cosh{\theta})]
 \tilde{q}_1+\tilde{w}\tilde{q}_2\sinh{\theta}-\frac{\tilde{w}^2}{2|E|}
 (\tilde{w}^2\tilde{q}_1^2-\tilde{q}_2^2)} \\
 & & \times [\mbox{sgn}(E)+2\Theta(-\eta_w\xi)\Theta(-E)\Theta(|\theta|-\ln{|w|})] \ ,
\end{eqnarray*}
and
\begin{eqnarray*}
 F_3(\bm{q}) &=& \! \sum_{\xi} \! \int^{\Lambda}_{-\Lambda} \! dE \! \int^{+\infty}_{-\infty} \! d\theta
 \frac{|E|(|w|\cosh{\theta}+\eta_E\eta_w\xi)^2}{[|w|(\xi w\cosh{\theta}+\eta_E)
 -(\eta_E\xi w+\cosh{\theta})]\tilde{q}_1-\tilde{w}\tilde{q}_2\sinh{\theta}+\frac{\tilde{w}^2}{2|E|}
 (\tilde{w}^2\tilde{q}_1^2-\tilde{q}_2^2)} \\
 & & \times [\mbox{sgn}(E)-2\Theta(-\eta_w\xi)\Theta(E)\Theta(|\theta|-\ln{|w|})] \\
 & & +\! \sum_{\xi} \! \int^{\Lambda}_{-\Lambda} \! dE \! \int^{+\infty}_{-\infty} \! d\theta
 \frac{|E|(|w|\cosh{\theta}-\eta_E\eta_w\xi)^2}{[|w|(\xi w\cosh{\theta}-\eta_E)
 -(\eta_E\xi w-\cosh{\theta})]\tilde{q}_1-\tilde{w}\tilde{q}_2\sinh{\theta}+\frac{\tilde{w}^2}{2|E|}
 (\tilde{w}^2\tilde{q}_1^2-\tilde{q}_2^2)} \\
 & & \times [\mbox{sgn}(E)-2\Theta(\eta_w\xi)\Theta(E)\Theta(|\theta|-\ln{|w|})] \ , \\
 F_4(\bm{q}) &=& \! \sum_{\xi} \! \int^{\Lambda}_{-\Lambda} \! dE \! \int^{+\infty}_{-\infty} \! d\theta
 \frac{|E|(|w|\cosh{\theta}+\eta_E\eta_w\xi)^2}{[|w|(\xi w\cosh{\theta}+\eta_E)
 +(\eta_E\xi w+\cosh{\theta})]\tilde{q}_1+\tilde{w}\tilde{q}_2\sinh{\theta}-\frac{\tilde{w}^2}{2|E|}
 (\tilde{w}^2\tilde{q}_1^2-\tilde{q}_2^2)} \\
 & & \times [\mbox{sgn}(E)+2\Theta(\eta_w\xi)\Theta(-E)\Theta(|\theta|-\ln{|w|})] \\
 & & +\! \sum_{\xi} \! \int^{\Lambda}_{-\Lambda} \! dE \! \int^{+\infty}_{-\infty} \! d\theta
 \frac{|E|(|w|\cosh{\theta}-\eta_E\eta_w\xi)^2}{[|w|(\xi w\cosh{\theta}-\eta_E)
 +(\eta_E\xi w-\cosh{\theta})]\tilde{q}_1+\tilde{w}\tilde{q}_2\sinh{\theta}-\frac{\tilde{w}^2}{2|E|}
 (\tilde{w}^2\tilde{q}_1^2-\tilde{q}_2^2)} \\
 & & \times [\mbox{sgn}(E)+2\Theta(-\eta_w\xi)\Theta(-E)\Theta(|\theta|-\ln{|w|})] \ ,
\end{eqnarray*}
where $\Lambda$ is the UV cutoff in energies.

An exact evaluation of $\Pi(0,\bm{q})$ is difficult. Instead of doing it, we shall determine it for
$v_1|q_1|,v_2|q_2|\ll D$. This can be done by calculating $\partial\Pi/\partial\Lambda$, yielding
\begin{equation}
 \frac{\partial}{\partial\Lambda}\Pi_{\Lambda}=-\frac{N}{8\pi^2v_1v_2\tilde{w}} \! \left[
 \frac{\partial}{\partial\Lambda}F_1(\bm{q})-\frac{\partial}{\partial\Lambda}F_2(\bm{q})\right] \!
 -\frac{N}{4\pi^2v_1v_2\tilde{w}^3} \! \left[\frac{\partial}{\partial\Lambda}F_3(\bm{q})
 +\frac{\partial}{\partial\Lambda}F_4(\bm{q})\right] , \label{wf2dd14}
\end{equation}
and
\begin{eqnarray*}
 \frac{\partial}{\partial\Lambda}F_{1/2}(\bm{q})=\mathcal{F}_{1/2}(\bm{q}) \ , ~~
 \frac{\partial}{\partial\Lambda}F_{3/4}(\bm{q})=\Lambda\mathcal{F}_{3/4}(\bm{q}) \ ,
\end{eqnarray*}
where $\Pi_{\Lambda}$ is the vacuum polarization with the energy cutoff $\Lambda$,
\begin{eqnarray*}
 \mathcal{F}_1(\bm{q}) &=& \! \sum_{\xi} \! \int^{+\infty}_{-\infty} \! d\theta
 \frac{[|w|(\xi w\cosh{\theta}+1)+(\xi w+\cosh{\theta})]\tilde{q}_1+\tilde{w}\tilde{q}_2\sinh{\theta}}
 {[|w|(\xi w\cosh{\theta}+1)-(\xi w+\cosh{\theta})]\tilde{q}_1-\tilde{w}\tilde{q}_2\sinh{\theta}}
 [1-2\Theta(-\eta_w\xi)\Theta(|\theta|-\ln{|w|})] \\
 & & +\! \sum_{\xi} \! \int^{+\infty}_{-\infty} \! d\theta\frac{[|w|(\xi w\cosh{\theta}-1)
 +(\xi w-\cosh{\theta})]\tilde{q}_1+\tilde{w}\tilde{q}_2\sinh{\theta}}{[|w|(\xi w\cosh{\theta}-1)
 -(\xi w-\cosh{\theta})]\tilde{q}_1-\tilde{w}\tilde{q}_2\sinh{\theta}}
 [1-2\Theta(\eta_w\xi)\Theta(|\theta|-\ln{|w|})] \\
 & & -\! \sum_{\xi} \! \int^{+\infty}_{-\infty} \! d\theta\frac{[|w|(\xi w\cosh{\theta}-1)
 -(\xi w-\cosh{\theta})]\tilde{q}_1+\tilde{w}\tilde{q}_2\sinh{\theta}}{[|w|(\xi w\cosh{\theta}-1)
 +(\xi w-\cosh{\theta})]\tilde{q}_1-\tilde{w}\tilde{q}_2\sinh{\theta}} \\
 & & -\! \sum_{\xi} \! \int^{+\infty}_{-\infty} \! d\theta\frac{[|w|(\xi w\cosh{\theta}+1)
 -(\xi w+\cosh{\theta})]\tilde{q}_1+\tilde{w}\tilde{q}_2\sinh{\theta}}{[|w|(\xi w\cosh{\theta}+1)
 +(\xi w+\cosh{\theta})]\tilde{q}_1-\tilde{w}\tilde{q}_2\sinh{\theta}} \ ,
\end{eqnarray*}
\begin{eqnarray*}
 \mathcal{F}_2(\bm{q}) &=& \! \sum_{\xi} \! \int^{+\infty}_{-\infty} \! d\theta
 \frac{[|w|(\xi w\cosh{\theta}+1)-(\xi w+\cosh{\theta})]\tilde{q}_1-\tilde{w}\tilde{q}_2\sinh{\theta}}
 {[|w|(\xi w\cosh{\theta}+1)+(\xi w+\cosh{\theta})]\tilde{q}_1+\tilde{w}\tilde{q}_2\sinh{\theta}} \\
 & & +\! \sum_{\xi} \! \int^{+\infty}_{-\infty} \! d\theta\frac{[|w|(\xi w\cosh{\theta}-1)
 -(\xi w-\cosh{\theta})]\tilde{q}_1-\tilde{w}\tilde{q}_2\sinh{\theta}}{[|w|(\xi w\cosh{\theta}-1)
 +(\xi w-\cosh{\theta})]\tilde{q}_1+\tilde{w}\tilde{q}_2\sinh{\theta}} \\
 & & +\! \sum_{\xi} \! \int^{+\infty}_{-\infty} \! d\theta\frac{[|w|(\xi w\cosh{\theta}-1)
 +(\xi w-\cosh{\theta})]\tilde{q}_1-\tilde{w}\tilde{q}_2\sinh{\theta}}{[|w|(\xi w\cosh{\theta}-1)
 -(\xi w-\cosh{\theta})]\tilde{q}_1+\tilde{w}\tilde{q}_2\sinh{\theta}}
 [-1+2\Theta(\eta_w\xi)\Theta(|\theta|-\ln{|w|})] \\
 & & +\! \sum_{\xi} \! \int^{+\infty}_{-\infty} \! d\theta\frac{[|w|(\xi w\cosh{\theta}+1)
 +(\xi w+\cosh{\theta})]\tilde{q}_1-\tilde{w}\tilde{q}_2\sinh{\theta}}{[|w|(\xi w\cosh{\theta}+1)
 -(\xi w+\cosh{\theta})]\tilde{q}_1+\tilde{w}\tilde{q}_2\sinh{\theta}}
 [-1+2\Theta(-\eta_w\xi)\Theta(|\theta|-\ln{|w|})] \ ,
\end{eqnarray*}
\begin{eqnarray*}
 \mathcal{F}_3(\bm{q}) &=& \! \sum_{\xi} \! \int^{+\infty}_{-\infty} \! d\theta
 \frac{(|w|\cosh{\theta}+\eta_w\xi)^2}{[|w|(\xi w\cosh{\theta}+1)-(\xi w+\cosh{\theta})]\tilde{q}_1
 -\tilde{w}\tilde{q}_2\sinh{\theta}}[1-2\Theta(-\eta_w\xi)\Theta(|\theta|-\ln{|w|})] \\
 & & +\! \sum_{\xi} \! \int^{+\infty}_{-\infty} \! d\theta\frac{(|w|\cosh{\theta}-\eta_w\xi)^2}
 {[|w|(\xi w\cosh{\theta}-1)-(\xi w-\cosh{\theta})]\tilde{q}_1-\tilde{w}\tilde{q}_2\sinh{\theta}}
 [1-2\Theta(\eta_w\xi)\Theta(|\theta|-\ln{|w|})] \\
 & & -\! \sum_{\xi} \! \int^{+\infty}_{-\infty} \! d\theta\frac{(|w|\cosh{\theta}-\eta_w\xi)^2}
 {[|w|(\xi w\cosh{\theta}-1)+(\xi w-\cosh{\theta})]\tilde{q}_1-\tilde{w}\tilde{q}_2\sinh{\theta}} \\
 & & -\! \sum_{\xi} \! \int^{+\infty}_{-\infty} \! d\theta\frac{(|w|\cosh{\theta}+\eta_w\xi)^2}
 {[|w|(\xi w\cosh{\theta}+1)+(\xi w+\cosh{\theta})]\tilde{q}_1-\tilde{w}\tilde{q}_2\sinh{\theta}} \ ,
\end{eqnarray*}
and
\begin{eqnarray*}
 \mathcal{F}_4(\bm{q}) &=& \! \sum_{\xi} \! \int^{+\infty}_{-\infty} \! d\theta
 \frac{(|w|\cosh{\theta}+\eta_w\xi)^2}{[|w|(\xi w\cosh{\theta}+1)+(\xi w+\cosh{\theta})]\tilde{q}_1
 +\tilde{w}\tilde{q}_2\sinh{\theta}} \\
 & & +\! \sum_{\xi} \! \int^{+\infty}_{-\infty} \! d\theta\frac{(|w|\cosh{\theta}-\eta_w\xi)^2}
 {[|w|(\xi w\cosh{\theta}-1)+(\xi w-\cosh{\theta})]\tilde{q}_1+\tilde{w}\tilde{q}_2\sinh{\theta}} \\
 & & +\! \sum_{\xi} \! \int^{+\infty}_{-\infty} \! d\theta\frac{(|w|\cosh{\theta}-\eta_w\xi)^2}
 {[|w|(\xi w\cosh{\theta}-1)-(\xi w-\cosh{\theta})]\tilde{q}_1+\tilde{w}\tilde{q}_2\sinh{\theta}}
 [-1+2\Theta(\eta_w\xi)\Theta(|\theta|-\ln{|w|})] \\
 & & +\! \sum_{\xi} \! \int^{+\infty}_{-\infty} \! d\theta\frac{(|w|\cosh{\theta}+\eta_w\xi)^2}
 {[|w|(\xi w\cosh{\theta}+1)-(\xi w+\cosh{\theta})]\tilde{q}_1+\tilde{w}\tilde{q}_2\sinh{\theta}}
 [-1+2\Theta(-\eta_w\xi)\Theta(|\theta|-\ln{|w|})] \ .
\end{eqnarray*}
\end{widetext}
In the above, we have assumed that $|\tilde{q}_1|,|\tilde{q}_2|\ll\Lambda$. Explication calculations
gives $\mathcal{F}_3(\bm{q})+\mathcal{F}_4(\bm{q})=0$ and
\begin{eqnarray*}
 \mathcal{F}_1(\bm{q}) &=& -\frac{8\tilde{w}^2[(w^2+1)\tilde{q}_1^2+\tilde{q}_2^2]}
 {(w^2+1)^2\tilde{q}_1^2-\tilde{w}^2\tilde{q}_2^2}\theta_{\Lambda} \\
 &=& -\mathcal{F}_2(\bm{q}) \ ,
\end{eqnarray*}
where $\theta_{\Lambda}$ is the cutoff in $\theta$, which is determined by the size of the first BZ.
Collecting the above results and integrating Eq. (\ref{wf2dd14}) from $\Lambda=0$ to $\Lambda=D$, we
obtain Eq. (\ref{wf2dd12}) with the nonuniversal constant $B=4\theta_{\Lambda}$.


\end{document}